\documentclass[preprint,12pt,authoryear]{elsarticle}
\usepackage{amssymb}

\usepackage[process=auto,crop=preview]{pstool}
\usepackage{booktabs}
\usepackage{amsmath}
\usepackage{hyperref}
\usepackage{filecontents}
\usepackage{bm}
\usepackage{multirow}
\usepackage{array}
\usepackage{makecell}

\journal{International Journal of Multiphase Flow}
\begin{document}

\begin{frontmatter}

\title{Best practices for velocity estimations in highly aerated flows with dual-tip phase-detection probes}

\author[mymainaddress1]{M. Kramer
\corref{mycorrespondingauthor}}
\cortext[mycorrespondingauthor]{Corresponding author}
\address[mymainaddress1]{UNSW Canberra, School of Engineering and Information Technology (SEIT), Canberra,
ACT 2610, Australia, m.kramer@adfa.edu.au, ORCID 0000-0001-5673-2751}

\author[mymainaddress2]{B. Hohermuth}
\address[mymainaddress2]{Laboratory of Hydraulics, Hydrology and Glaciology (VAW), ETH Zurich, Switzerland, hohermuth@vaw.baug.ethz.ch, ORCID 0000-0001-8218-0444}

\author[mymainaddress3]{D. Valero}
\address[mymainaddress3]{IHE Delft Institute for Water Education, Water Resources and Ecosystems, 2611 AX Delft, the Netherlands, d.valero@un-ihe.org, ORCID 0000-0002-7127-7547}

\author[mymainaddress4]{S. Felder}
\address[mymainaddress4]{Water Research Laboratory, School of Civil and Environmental Engineering, UNSW Sydney, NSW 2052, Australia, s.felder@unsw.edu.au, ORCID 0000-0003-1079-6658}

\begin{abstract}
Dual-tip phase-detection probes can be used to measure flow properties in gas-liquid flows. Traditionally, time-averaged interfacial velocities have been obtained through cross-correlation analysis of long time-series of phase fraction signals. Using small groups of detected particles, a recently developed adaptive window cross-correlation (AWCC) technique enables the computation of pseudo-instantaneous interfacial velocities and turbulence quantities in highly aerated flows, albeit subject to some smoothing which is due to the use of a finite window duration. This manuscript provides guidance on the selection of optimum processing parameters for the AWCC technique, additionally addressing shortcomings such as velocity bias correction in turbulent flows and extrapolation of turbulence levels to single particles. The presented technique was tested for three highly turbulent air-water flows: smooth and rough-wall boundary layers (tunnel chute and stepped spillway), as well as breaking shear layer flows of a hydraulic jump. Robust estimations of mean velocities and velocity fluctuations were obtained for all flow situations, either using dual-tip conductivity or fiber optical probe data. 
The computation of integral time scales and velocity spectra is currently limited by the data rate and must be treated with caution. 
\end{abstract}

\end{frontmatter}

\section{Introduction}\label{sec:intro}

\textit{White waters} can be observed in free-surface flows such as hydraulic jumps or high-speed flows over smooth or rough beds. Aerated flows present a complex structure \citep{Wilhelms2005bubbles} and its inception is strongly linked to the presence of turbulence close to the free-surface \citep{Brocchini01JFM, Valero18Reformulating}. Entrained air limits the use of common monophase flow measurement methods and has motivated the development of specialised air-water flow instrumentation \citep{Jones1976, Welle1985, Kataoka1986, Revankar1993, Crowe2011}.
 
\begin{figure}[h!]
\includegraphics[width=0.9\textwidth]{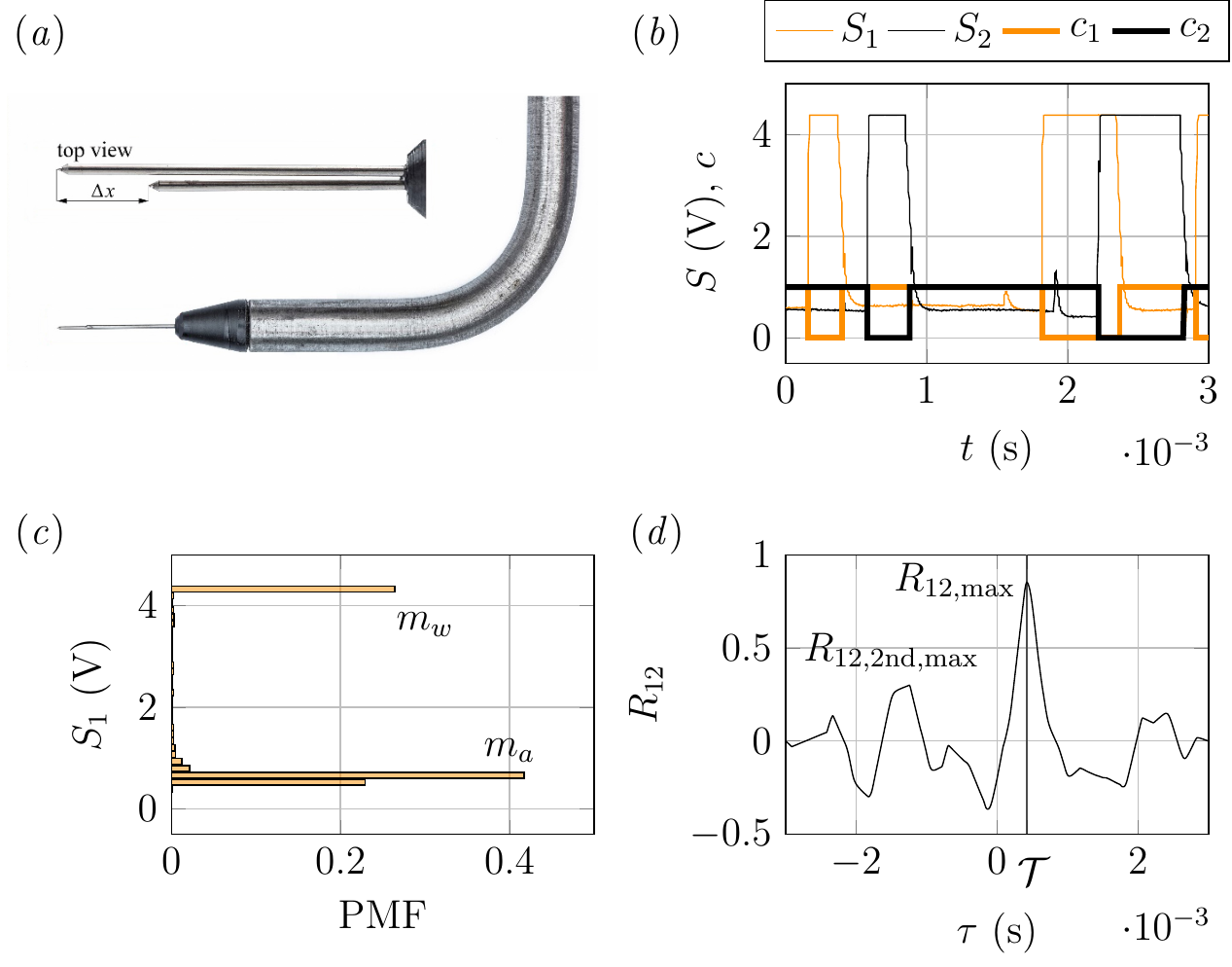}
\caption{Phase-detection probe design and recorded signals: (\textit{a}) dual-tip phase-detection conductivity probe (CP) (\textit{b}) typical leading and trailing tip raw voltage signal ($S_1$, $S_2$) and instantaneous void fraction ($c_1$, $c_2$); tunnel chute data with $C$ $\approx$ 0.6 (\textit{c}) probability mass function (PMF) of the raw voltage signal with $m_w$, $m_a$ = mode of the water and air phase, respectively (\textit{d}) cross-correlation function $R_{12}$ versus time lag $\tau$,  with $R_{12,\mathrm{max}}$ = 0.85 and secondary peak ratio SPR = $R_{12,2\mathrm{nd},\mathrm{max}}$/$R_{12,\mathrm{max}}$ $\approx$ 0.45; $\mathcal{T}$ = interfacial travel time.}
\label{fig:probe}
\end{figure}

Dual-tip phase-detection probes (figure \ref{fig:probe}\textit{a}) have been used to measure air-water flow properties intrusively since the early works of \cite{Neal1963} and \cite{Herringe1976}. A single-threshold technique \citep{Cartellier1991,Felder2015} is typically used to analyse the raw voltage signal of the probe $S(\bm{x},t)$ (figure \ref{fig:probe}\textit{b,c}), providing an ideal square wave time series of instantaneous void fractions, with \mbox{$c(\bm{x},t)$ = 1} for air and \mbox{$c(\bm{x},t)$ = 0} for water. From the instantaneous void fraction signal, time-averaged void fraction $ C (\bm{x}) $, time-averaged particle count rate $F (\bm{x}) $ and chord times can be deducted  \citep{Chanson2016}. Note  that $\bm{x}$ refers to the spatial coordinate of the leading tip of the probe and $t$ refers to the time coordinate. For the sake of conciseness, $\bm{x}$ is dropped from the following analysis, although flow dependent variables can naturally vary across space.

\subsection{Conventional velocity and turbulence level estimations}
\label{sec:conventional}
Velocity fluctuations in bubbly flows were investigated by \cite{Sene84, Cartellier1992, Cartellier1998, Serdula1998}, using the phase function gradients of a single-tip fiber optical (FO) signal. Similarly, \cite{Chang2003} and \cite{Zhang14} used event detection techniques to measure bubble velocities with a single-tip fiber optical probe. While these techniques are applicable for finely dispersed flows, dual-tip probes are better suited to cover a wide range of void fractions \citep{Chanson2002,Chanson2016, Valero2018ANN}, which can be common in free-surface air-water flows.

The estimation of the local time-averaged interfacial velocities with dual-tip phase-detection probes is based upon the cross-correlation analysis of the simultaneously sampled signals of the two probe tips. 

The cross-correlation can provide the time shift between the probe tip signals \citep{Cain1981, Chanson2002, Crowe2011}, i.e. the most probable time lag ($\mathcal{T}$) between a pair of phase functions. Knowing the streamwise distance between the probe tips ($\Delta \, x$), the expected value of the mean interfacial velocity can be obtained as:
\begin{equation}
\langle U \rangle 
\approx \frac{\Delta\, x}{\mathcal{T}}
\label{eq:classic-xcorr}
\end{equation}

Here the operator $\langle...\rangle$ denotes time-averaging. \cite{Chanson2002a} suggested that indirect turbulence estimations could be obtained on the basis of the standard deviation of the auto- and cross-correlation functions ($\sigma_{11}$ and $\sigma_{12}$) of the raw phase-detection probe signal ($S_1$, $S_2$):
\begin{equation}
Tu = u_{rms}/\langle U \rangle \approx \sqrt{\sigma_{12}^2-\sigma_{11}^2}/\mathcal{T}
\label{eg:Tuold}
\end{equation}
Based on eq. (\ref{eg:Tuold}), any imperfections that imply a decorrelation between $S_1$ and $S_2$ can result in a higher turbulence level, which may not be related to the flow properties but to experimental defects. Apart from actual turbulent velocity fluctuations, several effects could cause a reduction in the correlation of the two signals, including 1) sampling rate: only an infinite sampling rate would yield a perfect correlation 2) transversal tip separation: the transverse separation leads to different chord times of the leading and trailing tips, resulting in a loss of correlation 3) probe misalignment 4) particle-probe interactions: the deformation of an air or water particle during the piercing of a probe tip leads to a lower correlation because the trailing tip detects a differently shaped particle.

\subsection{Adaptive window cross-correlation (AWCC) technique}
Recently, \cite{Kramer19AWCC} developed an adaptive window cross-correlation (AWCC) technique for processing dual-tip phase-detection probe signals. The AWCC Matlab source code is available on GitHub \citep{AWCC}. The main innovation of the AWCC technique is the segmentation of dual-tip phase-detection probe signals into very short windows (such as those in figure \ref{fig:probe}\textit{b}) and the subsequent estimation of a pseudo-instantaneous velocity for each window. The window duration is determined upon a defined number of dispersed phase particles ($N_p$) contained in a given segment. Each particle is defined as a pair of interfaces; for instance, the leading and the trailing tip signals shown in figure \ref{fig:probe}\textit{b} contain two dispersed phase particles each.

The segmentation of the signal into a series of windows provides a time series of pseudo-instantaneous interfacial velocities ($U(t)$) in highly aerated flows, enabling a direct computation of turbulence parameters. Here, the term \textit{pseudo}-instantaneous is used because each velocity estimation is representative of a number of $N_p$ particles, i.e. it is averaged over the window duration. It is expected that instantaneous velocity estimations can be obtained on the limit $N_p \rightarrow 1$. The AWCC was validated by stochastic modelling, showing that the method is accurate, although it was noticed that the uncertainty (e.g. velocity bias) may increase with increasing velocity fluctuations \citep{Kramer19AWCC}.

Experimentally recorded signals are different from synthetic signals and the utilization of the AWCC technique in laboratory or large-scale applications requires an appropriate selection of processing parameters, including the number of particles $N_p$ and threshold values for the maximum cross-correlation coefficient and the secondary peak ratio  \citep{Kramer19AWCC} (figure \ref{fig:probe}\textit{d}). To facilitate the selection of processing parameters, this work explores the sensitivity of the AWCC technique through a re-analysis of previous phase-detection probe measurements \citep{Felder2017,Felder2019,Montano19}. The AWCC technique is detailed in section \ref{sec:signalprocessing}, followed by a description of the selected datasets (section \ref{sec:datasets}) and a presentation of the sensitivity analysis (section \ref{sec:results}). In section \ref{sec:discussion}, the AWCC technique is compared to conventional signal processing approaches and best practices are discussed.

\subsection{A glimpse on measurement accuracy}
The performance of  phase-detection probes has been addressed by several researchers (table \ref{tab:uncertainties}). A conclusive statement on the measurement uncertainty is challenging, given the variety of probe designs and flow conditions. Table \ref{tab:uncertainties} summarises previous literature on the accuracy of void fraction $C$ and interfacial velocity $\langle U \rangle$.

\begin{table}[h!]
\caption{Reported measurement uncertainties for  phase-detection probe data; CP: phase-detection conductivity probe; FO: phase-detection fiber optical probe.}
\begin{footnotesize}
\centering
\begin{tabular}{l c c c c c c c}
\toprule
Reference & $\Delta C/C$ & $\Delta \langle U \rangle / \langle U \rangle$ & comment & probe \\
\midrule
\cite{Cummings97} & $\approx \pm 0.001/C$ &  $\pm$ 10$\%$ & $ C < 0.05$ & dual-tip CP \\
\cite{Cummings97}& $\pm$ 2$\%$ & $\pm$ 5$\%$ & $0.05 < C < 0.95$ & dual-tip CP \\
\cite{Cummings97}& $\approx \pm 0.001/(1-C)$ & $\pm$ 10$\%$ & $ C > 0.95$ &dual-tip CP \\
\cite{Barrau99} & 0$\%$ to $-$16$\%$ &  & bubbly flow &dual-tip FO \\
\cite{Boes2000} & $< -5\%$ & $< \pm 5\%$ & highly aerated flow &dual-tip FO \\
\cite{Kiambi03} & $-$6$\%$ to $-$14$\%$ & $<+5\%$ & bubbly flow &dual-tip FO \\
\cite{Cartellier2006} & $<-5\%$ & $< 5\%$ &  thin and sharp tip &dual-tip FO  \\
\cite{Vejrazka2010} & $-$10$\%$ &  & \thead{bubbly flow \\ correction applied} &single-tip FO \\  
\bottomrule
\end{tabular}
\end{footnotesize}
\label{tab:uncertainties}
\end{table}
\newpage
The phase-detection probes of the studies re-analysed in this manuscript compared well with other conductivity probes \citep[e.g.][]{Cummings97} and fiber optical probes \citep[e.g.][]{Boes2000}, as shown in \cite{Felder2017,Felder2019}. The measurement uncertainty of the conductivity probes in this manuscript is  therefore expected to be similar to the data presented in table \ref{tab:uncertainties}. Additional uncertainty may be linked to sampling parameters as shown for the sampling duration in \ref{app:convergence}.

\section{Signal processing}
\label{sec:signalprocessing}

\begin{figure}[h!]
\centering
\includegraphics[width=1\textwidth]{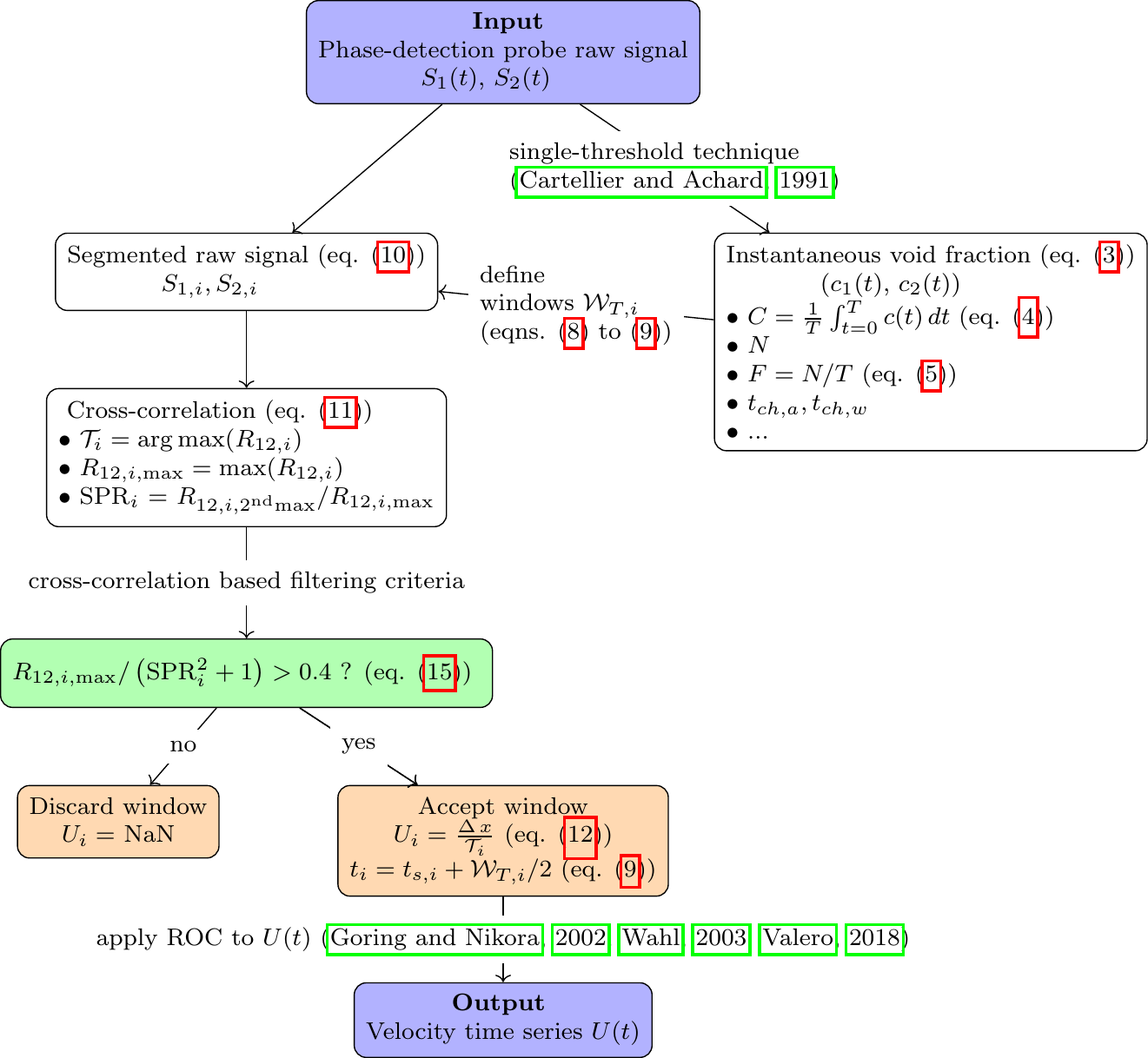}
\caption{Flowchart of the AWCC technique for dual-tip phase-detection probe data; ROC: Robust Outlier Cutoff; NaN: Not a Number.}
\label{fig:flowchart}
\end{figure}

The processing steps of the AWCC technique involve the calculation of basic multiphase flow parameters (void fraction, chord times, particle count rate), segmentation of the signal into adpative windows, filtering and  calculation of pseudo-instantaneous interfacial velocities (figure \ref{fig:flowchart}). Further analyses on the obtained velocity time series are presented in section \ref{sec:timeseries}.

\subsection{Void fraction and particle count rate}
The raw signals of the leading ($S_1$) or trailing tip ($S_2$) of a phase-detection probe often differ from an ideal square-wave signal because of wetting and drying processes and electrical noise (figure \ref{fig:probe}\textit{b}). Therefore,  single-threshold \citep{Cartellier1991,Toombes2002,Felder2015} or double-threshold techniques \citep{Welle1985,Wang17} are commonly used to obtain the instantaneous void fraction $c(t)$. For the leading tip signal, a single-threshold criteria based on 50 \% of the intermodal voltage range can be written as:

\begin{equation}
\label{eq:singlethreshold}
\frac{S_1(t) - m_a}{(m_w - m_a )} \, \begin{cases}
 \geq 0.5 \rightarrow c_1(t) = 0 \\
< 0.5 \rightarrow c_1(t) = 1
\end{cases}
\end{equation}
where $m_w$ and $m_a$ are the modes of the probability mass function of the raw signal (figure \ref{fig:probe}\textit{c}), corresponding to the water and air phases respectively. The time averaged void fraction $C$ is calculated as:
\begin{equation}
\label{eq:voidfraction}
 C = \frac{1}{T}\int_{t = 0}^{T} c_1(t) \, dt 
\end{equation}
where the sampling duration $T$ is significantly larger than the time scale of turbulent processes.

The instantaneous void fraction can provide air and water chord times ($t_{ch,a}$ and $t_{ch,w}$), the total number of particles in the signal $N$ (equal to half the number of detected interfaces) and the time-averaged particle count rate can be estimated as:
\begin{equation}
\label{eq:particlecount}
 F = \frac{N}{T}
\end{equation}

\subsection{Signal segmentation}
The adaptive window duration $\mathcal{W}_T$ is determined to enclose a certain number of individual bubble-droplet events \citep{Kramer19AWCC}. Given the air and water chord times, the start and end times for the $i^{th}$ window are determined on the basis of the leading tip signal $(S_1)$ and a selected number of particles $N_p$. The start time of the first window  is $t_{s,1}=0$ and the start times for $2 \leq i \leq n_{\mathcal{W}}$  are computed as:
\begin{equation}
t_{s,i} =   \sum_{j = 1}^{N_p  \cdot (i-1)}t_{ch,a,j} + \sum_{j = 1}^{N_p  \cdot (i-1)}t_{ch,w,j}  \quad \{ i \in \mathbb{Z}^{+}: 2 \leq i \leq n_{\mathcal{W}} \}
\label{eq:windows1a}
\end{equation}
where $n_{\mathcal{W}}= \lfloor N/N_p \rfloor$ is the number of windows and $\lfloor \cdot \rfloor$ is the floor function. The end times read: 
\begin{equation}
t_{e,i} =   \sum_{j = 1}^{N_p  \cdot i}t_{ch,a,j} + \sum_{j = 1}^{N_p  \cdot i}t_{ch,w,j} \quad \{ i \in \mathbb{Z}^{+}: 1 \leq i \leq n_{\mathcal{W}} \}
\label{eq:windows1b}
\end{equation}

 Equations (\ref{eq:windows1a}) and (\ref{eq:windows1b}) imply that each segment of the leading tip signal ($S_1$) contains  $N_p$ particles, whereas the number of particles in the trailing tip signal ($S_2$) can be different. The window duration for the $i^{th}$ window is:
\begin{equation}
\label{eq:windows1}
\mathcal{W}_{T,i} = t_{e,i}  - t_{s,i} 
\end{equation}

It is further assumed that the time $t_i$ of the velocity estimation is  represented by the midpoint:
\begin{equation}
\label{eq:windows2}
t_i = (t_{s,i} + t_{e,i})/2
\end{equation}
and the signals contained in the $i^{th}$ window are: 
\begin{equation}
\label{eq:segment}
S_{1,i} = S_1(t_{s,i} \leq t \leq  t_{e,i} ) , \quad 
S_{2,i} = S_2(t_{s,i} \leq t \leq  t_{e,i} )
\end{equation}

\subsection{Velocity estimation and filtering}
The interfacial travel time for an arbitrary window is obtained by means of the cross-correlation of the leading and trailing tip signals:
{\footnotesize
\begin{equation}
 R_{12, i}(\tau) 
  = 
 \frac{
 \sum_{t=t_{s,i}}^{t_{s,i}+\mathcal{W}_{T, i}}
 \Big(	
 \left[
 S_{1}(t)-\langle S_{1,i} \rangle \right]
 \cdot
 \left[ S_{2}(t+\tau)- \langle S_{2,i} \, \rangle
 \right]
 \Big)
 }
 {
 \sqrt{\sum_{t=t_{s,i}}^{t_{s,i}+\mathcal{W}_{T, i}}
 \Big(
 S_{1}(t)-\langle S_{1,i} \rangle
 \Big)^2}
 \cdot
 \sqrt{\sum_{t=t_{s,i}}^{t_{s,i}+\mathcal{W}_{T, i}}
 \Big(S_{2}(t+\tau)-\langle S_{2,i} \rangle
 \Big)^2 }
 } 
\label{eq:xcor_basic}
\end{equation}}
where $\tau$ is the time lag and $R_{12, i}$ the cross-correlation function. The peak of $R_{12,i}$ indicates the interfacial travel time $\mathcal{T}_i$ ($ =\arg \max(R_{12, i})$) for which both signals are best correlated (figure \ref{fig:probe}\textit{d}). A longitudinal pseudo-instantaneous velocity is estimated through eq. (\ref{eq:classic-xcorr}) for each window as:
\begin{equation}
\label{eq:velocity}
\Big[ \, \langle U  \rangle \, \Big]_{t_{s,i}} ^{t_{s,i} + \mathcal{W}_{T,i}}
\approx U_i = \frac{\Delta\, x}{\mathcal{T}_i}
\end{equation}

Shortcomings can be caused by oblique particle impacts, resulting in large velocities due to a nearly simultaneous recording of interfaces at both tips \citep{Thorwarth2008}, or by windows capturing different particle events; while the cross-correlation functions appear reasonable, the interfacial velocity could be overestimated.

To limit erroneous velocity estimations, a robust filtering approach was proposed by \cite{Kramer19AWCC}, comprising thresholds for 1) the maximum cross-correlation coefficient $R_{12,i,\mathrm{max}}$, representing the similarity between $S_{1,i}$ and $S_{2,i}$ and 2) the secondary peak ratio SPR$_i$ = $R_{12,i,2\mathrm{nd},\mathrm{max}}$/$R_{12,i,\mathrm{max}}$. SPR$_i$ is defined as the ratio of the second tallest peak to the first tallest peak of the cross-correlation function (figure \ref{fig:probe}\textit{d}) and indicates the presence of one or more events within one windowNote that both parameters detect different physical processes with the aim to characterise the \textit{quality} of a cross-correlation function. The choice of threshold values for $R_{12,i,\mathrm{max}}$ and SPR$_i$ as well as a combined filtering criteria are discussed in section \ref{sec:filtering}.

In addition to the cross-correlation based filtering, a simplified version of the despiking method proposed by \cite{Goring02}, as modified by \cite{Wahl03}, is applied to the estimated velocity time series $U_i$ at $t_i $. The Robust Outlier Cutoff \citep[ROC, see][]{Valero2018} is performed iteratively until no more outliers are rejected. Outliers are accounted as NaN data.

\subsection{Time series analysis}
\label{sec:timeseries}
Similar to time series obtained with laser Doppler anemometers (LDA), phase-detection probe velocity measurements $U(t)$ may overestimate the mean velocity because more particles pass the measurement volume/hit the probe tips at higher velocities \citep{Britz1996,Kramer19AWCC}. This shift of the mean velocity towards a higher value is known as velocity bias \citep{McLaughlin1973}. Herein, the following correction was used to eliminate the velocity bias for phase-detection probe measurements:
 
\begin{equation}
\label{eq:meanvelocity}
\langle U \rangle = 
\frac{\sum_{i=1}^{n_{\mathcal{W}}} w_i U_i }{\sum_{i=1}^{n_{\mathcal{W}}} w_i}
\end{equation}

\begin{equation}
\label{eq:rms}
 u_{rms} =  \sqrt{
 \frac{ 
 \sum_{i=1}^{n_{\mathcal{W}}} \Big( U_i - \langle 
 U 
 \rangle
 \Big)^2 w_i}{\sum_{i=1}^{n_{\mathcal{W}}} w_i} }
\end{equation}
where $w_i$ is a weighting factor and the summations are taken over the entire ensemble. Common weighting schemes for LDA measurements comprise inverse absolute velocity weighting \citep{McLaughlin1973}, particle interarrival time weighting  \citep{Hoesel77} and particle residence time weighting \citep{Buchhave79}, where the latter is the most appropriate method to eliminate velocity bias in LDA data \citep{Velte2014}.

As residence times are not available for air-water flow data, the synthetic signals of \cite{Kramer19AWCC} were re-analysed and three different weighting schemes for the AWCC technique were tested. In \cite{Kramer19AWCC}, a virtual dual-tip phase-detection probe was immersed in a pattern of randomly distributed fluid particles (following a gamma distribution with 3 mm mean diameter), simulating 1D turbulent motion. The tip separation was $\Delta x$ = 4 mm and the probe was sampled for $T = 10$ s at $f$ = 20 kHz, generating instantaneous void fraction signals. Additional information on synthetic signals and the stochastic particle generator can be found in \cite{Valero19SBG,Kramer19Particle}.

\begin{table}[h!]
\caption{Comparison of different weighting schemes using synthetic signals; turbulence level $Tu = u_{rms}/\langle U \rangle$; integral time scale $T_{uu} = 0.06$ s for all flow conditions; data processed with $N_p$ = 2 and errors calculated as $e_{\langle U \rangle} = \left(\langle U \rangle - \langle U \rangle_{\mathrm{sim}}\right)/\langle U \rangle_{\mathrm{sim}}$.}
\begin{footnotesize}
\centering
\begin{tabular}{cccc | cc | cc | cc | cc }
\toprule
\multicolumn{4} {c|} {Simulation} & \multicolumn{8} {c} {Relative errors using different weighting schemes} \\
$\langle U \rangle_{\mathrm{sim}}$ & $Tu_{\mathrm{sim}}$ &  $C_{\mathrm{sim}}$ & $F$  & $e_{\langle U \rangle}$ & $e_{Tu}$ & $e_{\langle U \rangle}$ & $e_{Tu}$ & $e_{\langle U \rangle}$ & $e_{Tu}$ & $e_{\langle U \rangle}$ & $e_{Tu}$\\
$ $(m/s) & (-) &  (-) & (1/s) & ($\%$) & ($\%$) & ($\%$) & ($\%$) & ($\%$) & ($\%$) & ($\%$) & ($\%$) \\
\multicolumn{4} {c|} {} & \multicolumn{2}{c|}{no weighting} & \multicolumn{2} {c|} {$1/\lvert U_i \rvert$} & \multicolumn{2} {c|} {$t_i-t_{i-1}$} & \multicolumn{2} {c} {$\mathcal{W}_{T,i}$}\\
\hline
3.0 & 0.1  & 0.3 & 142  & 0.9 & -1.9  & 0.4 & 0.8  & 0.6 & 1.0 & 0.6 & 1.0 \\
3.0 & 0.1  & 0.1 & 56   & 1.4 & -2.3 &1.5 & 0.0 & 1.8 & -3.2 &1.8 & -5.6 \\
3.0 & 0.2  &  0.3 & 140   & 3.0 & -9.5  & 0.3 & -4.1 & 1.2 & -5.5 & 1.2 & -6.3\\
3.0 & 0.2  &  0.1 & 57    & 3.8 & -10.0 &  1.1 & -8.5 &1.8 & -12.4 & 1.5 & -7.9\\
3.0 & 0.3  & 0.3 & 144   & 7.6 & -11.7  & 0.3 & -0.9 & 2.0 & -5.5 & 1.1 & -5.2\\
3.0 & 0.4  & 0.3 & 131   & 19.0 & -23.9  & 3.7 & -10.6 & 6.2 & -15.5 & 4.5 & -14.0\\
3.0 & 0.5 & 0.3 & 143   & 28.7 & -33.2  & 7.5 & -15.9 & 11.9 & -22.7 & 9.0 & -19.3\\
\bottomrule
\end{tabular}
\end{footnotesize}
\label{tab:weighting}   
\end{table}

To determine the best weighting scheme for the AWCC technique, the synthetic void fraction signals were re-analysed using 1) inverse absolute velocity weighting ($w_i = 1/\lvert U_i \rvert$), 2) interarrival time weighting ($w_i = t_i - t_{i-1}$) and 3) window duration weighting ($w_i = \mathcal{W}_{T,i}$). The results were compared with arithmetic statistics ($w_i =1$, no weighting), highlighting that all weighting schemes reduced the errors in terms of mean velocities and velocity fluctuations (table \ref{tab:weighting}). 
Despite having the lowest errors for the investigated 1D synthetic signals, inverse velocity weighting led to unreasonable mean velocities in free shear regions (e.g. in hydraulic jumps), possibly caused by the assignment of very large weights to near-zero velocities. Interarrival time weighting and window duration weighting performed comparably. Hereafter, mean velocities $\langle U \rangle$ and $rms$ velocity fluctuations $u_{rms}$ were evaluated using equations (\ref{eq:meanvelocity}) and (\ref{eq:rms}) with $w_i = \mathcal{W}_{T,i}$ (window duration weighting).

Further quantities of interest include turbulent properties such as the auto-correlation function (ACF) and the power spectral density (PSD). When calculating these properties from AWCC (likewise LDA) velocity time series, the effect of non-uniform sampling must be considered. \cite{Benedict2000} and \cite{Damaschke2018} compared  state-of-the-art ACF and PSD estimation methods for LDA data, for example slotting and interpolation techniques. In the present study, nearest-neighbor (NN) resampling and autoregressive moving-average (ARMA) model fitting \citep{Broersen2000} was used for the AWCC technique to estimate ACF and PSD. NN resampling was performed at twice the mean data rate ($f_m$) of the original time series, where $f_m$ is defined as the amount of valid velocity information per time. This method allows spectral estimates up to $0.5f_{m}$.

\section{Datasets and experimental facilities}
\label{sec:datasets}
\subsection{General remarks}
\begin{figure}[!htb]
\minipage{0.295\textwidth}
  \includegraphics[width=\linewidth]{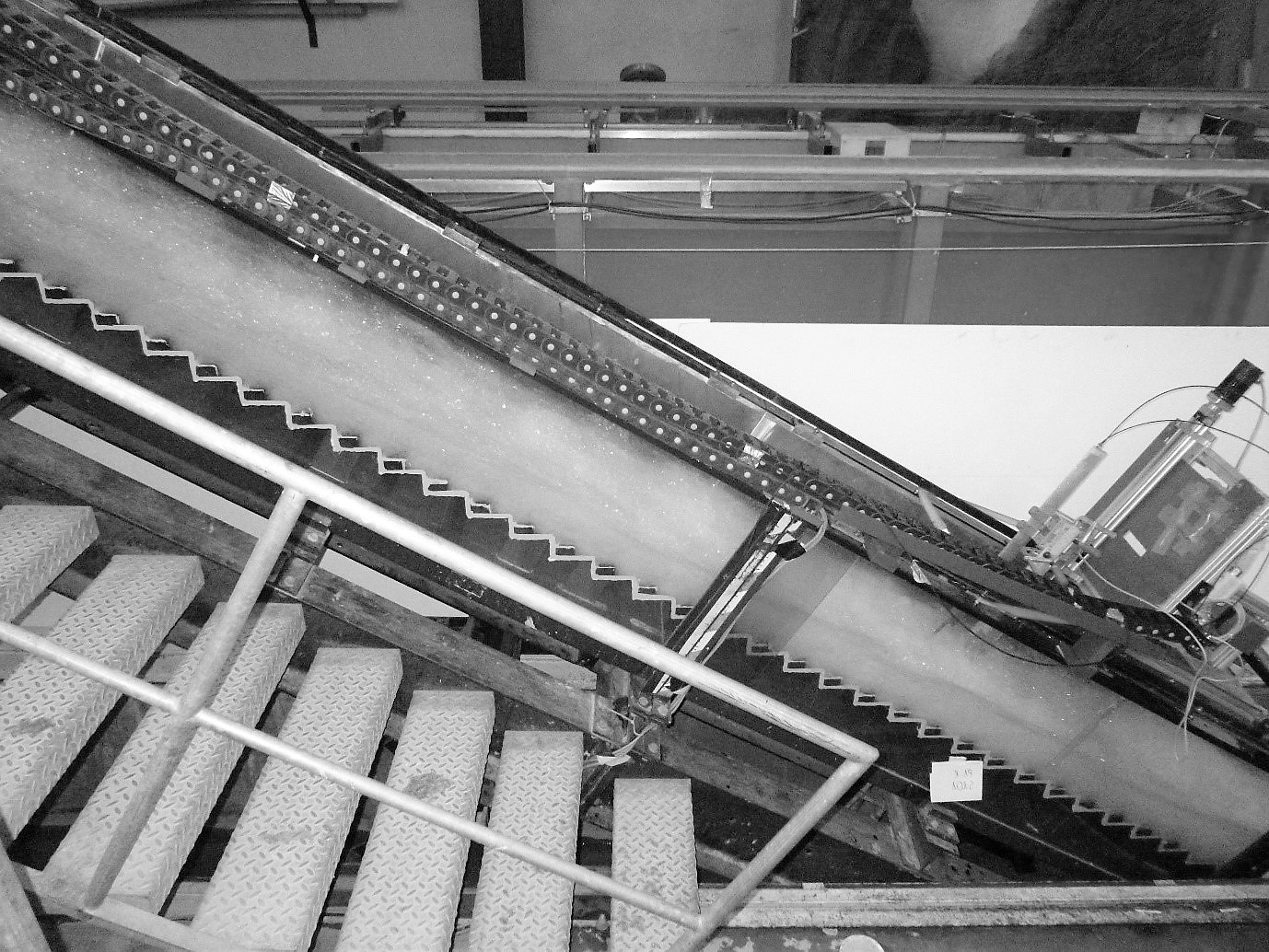}
\endminipage\hfill
\minipage{0.33\textwidth}
 \includegraphics[width=\linewidth]{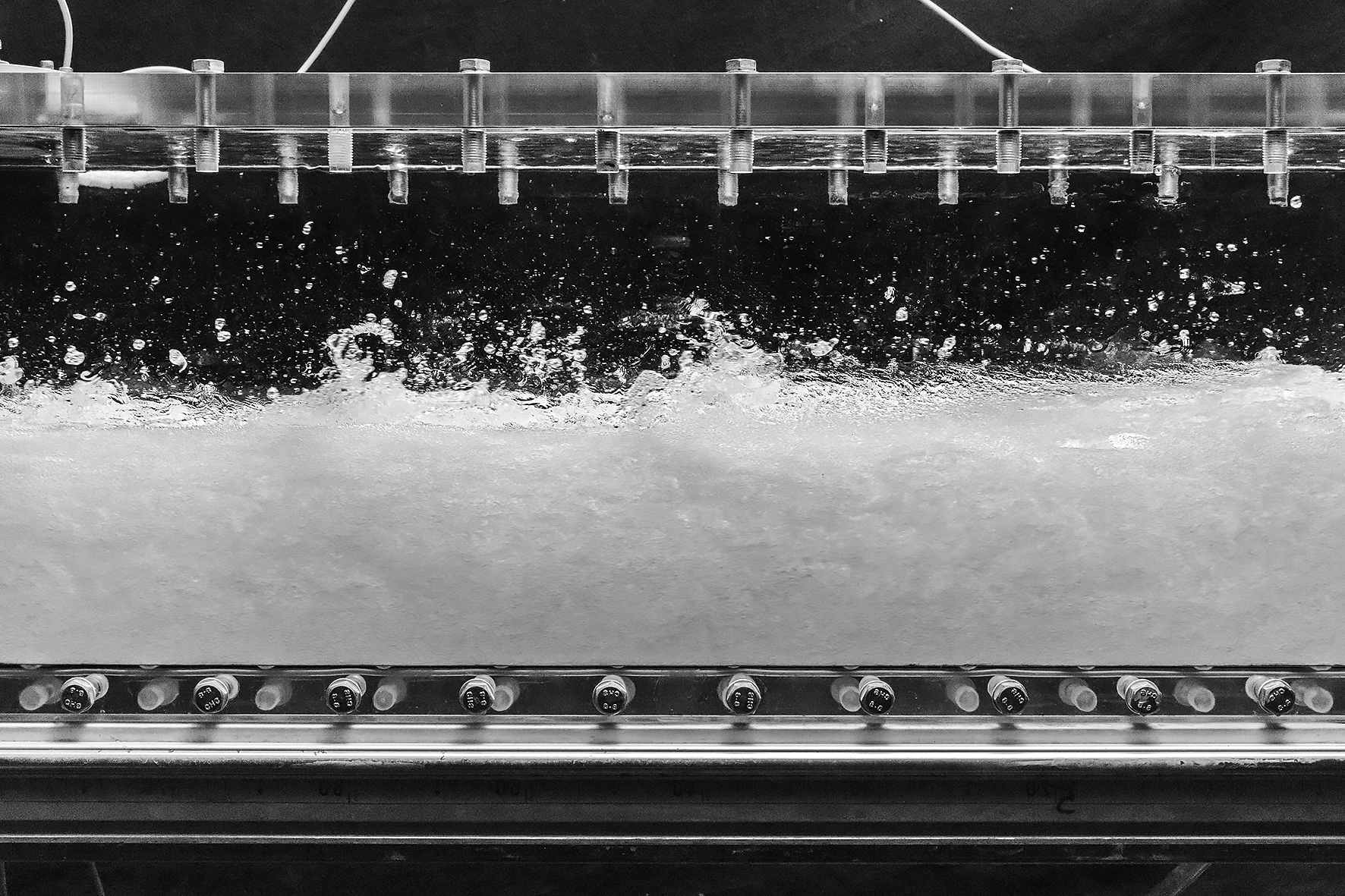}
\endminipage\hfill
\minipage{0.32\textwidth}%
 \includegraphics[width=\linewidth]{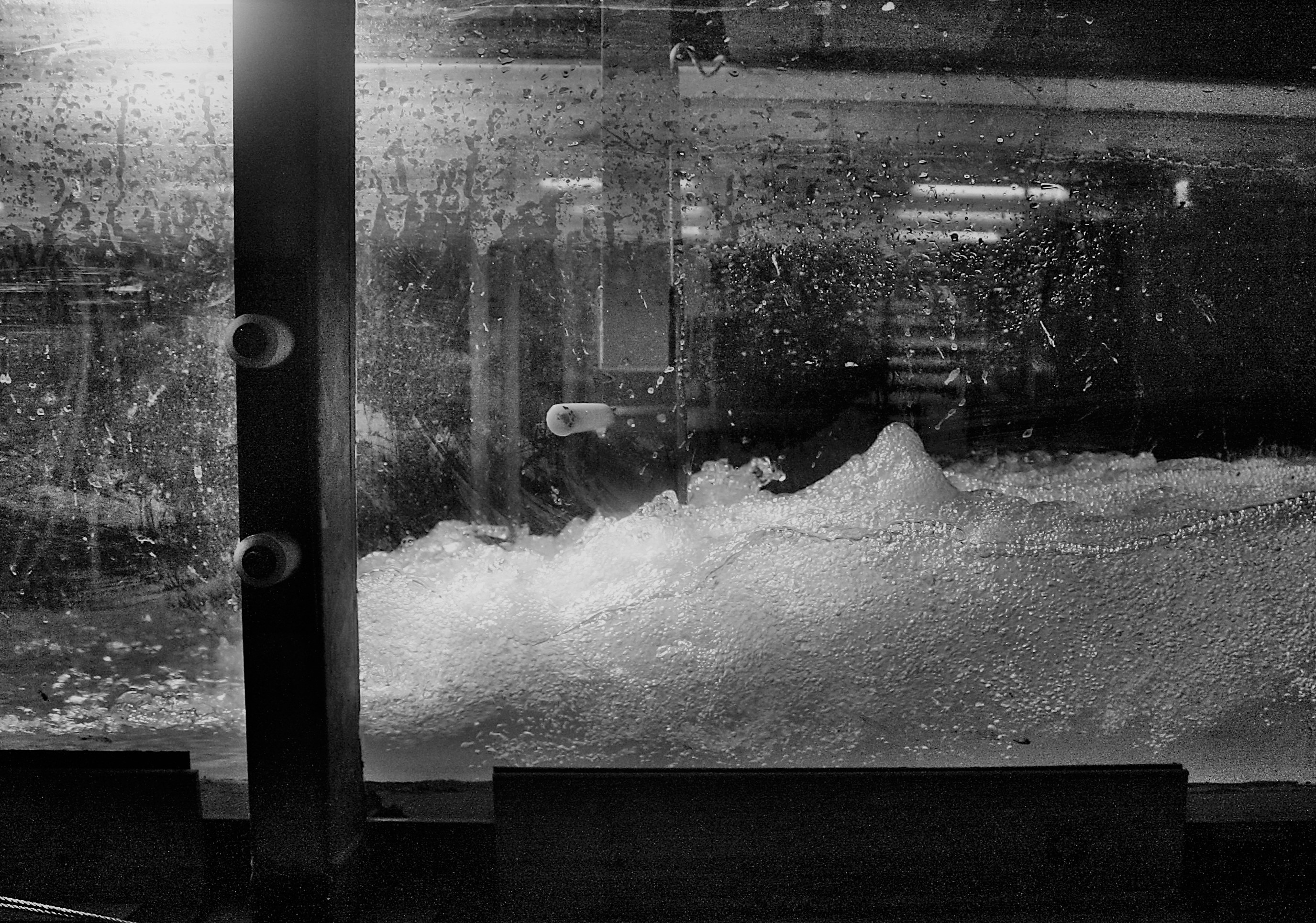}
\endminipage
\label{fig:setup}
\caption{Case studies; flow direction from left to right (\textit{a}, left) high-velocity flows in a stepped spillway; $d_c/h$ = 9.5; $Re$ = 1.2$\cdot$10$^5$ (\textit{b}, middle) high-velocity flows in a tunnel chute; $H$ = 15 m; $h_g/h_{\mathrm{us}}$ = 0.4; $Re$ = 1.1$\cdot$10$^6$ (\textit{c}, right) fully aerated hydraulic jump; $d_1$ = 0.03 m; $Fr$ = 5.1; $Re$ = 8.3$\cdot$10$^4$; photography courtesy of Laura Montano.}
\end{figure}

Three datasets comprising high-velocity air-water flows down a stepped spillway (figure \ref{fig:setup}\textit{a}), high-velocity air-water flows in a smooth tunnel chute (figure \ref{fig:setup}\textit{b}) and a fully aerated hydraulic jump (figure \ref{fig:setup}\textit{c}) were re-analysed with the AWCC. In all datasets, dual-tip phase-detection intrusive probes were used, comprising conductivity probes (CP) manufactured in the Water Research Laboratory (WRL, UNSW Sydney) (figure \ref{fig:probe}\textit{a}) and a commercial fiber optical (FO) probe from RBI. 

All dual-tip probes had a side-by-side design, as recommended by \cite{Felder2019}. The conductivity probes were similar, with inner electrodes of platinum wire \mbox{($\phi$ = 0.125 mm)} and outer electrodes of hypodermic needles \mbox{($\phi$ = 0.6 mm);} both electrodes were insulated using epoxy. Every time an air entity is pierced by a probe tip, the voltage signal drops to about 0.5 V and when a probe tip is in water, the voltage signal increases to about 4.1 V (figure \ref{fig:probe}\textit{b}). The fiber optical probe tips record the air-water phases via a change in light refraction in front of the needle tips. For each dataset, the longitudinal separation distance between the leading and trailing tips $\Delta \, x$, the sampling time $T$, the sampling rate $f$ and the Reynolds number are provided in table \ref{tab:experimentalconditions}. The choice of the sampling parameters fulfilled the requirements set by earlier sensitivity analyses \citep{Felder2015}. A convergence analysis of the stepped spillway data showed that sampling durations $T> 45$ s could further improve the measurement accuracy (\ref{app:convergence}). The raw voltage signals were recorded with the same LabVIEW-based data acquisition system using a high-speed acquisition unit (National Instruments USB-6366) and further details on the overall probe designs can be found in \cite{Felder2017} and \cite{Felder2019}. 

\begin{table}[h!]
\caption{Experimental flow condition and dual-tip phase-detection probe specifications for the re-analysed datasets; CP: phase-detection conductivity probe; FO: phase-detection fiber optical probe;  Reynolds number $Re$ = $q/\nu$, with $\nu$ the kinematic viscosity of water and $q$ the specific discharge; $f$: sampling rate, $T$: sampling duration; $\Delta \, x$: streamwise tip separation distance.
}
\begin{footnotesize}
\centering
\begin{tabular}{l c c c c c c c}
\toprule
Reference & $q$ & $Re$ & flow type & probe& $f$  & $T$  & $\Delta \, x$ \\
(-) &  (m$^2$/s) & (-) & (-) & (-) & (kHz)  & (s)  & (mm)\\
\midrule
\cite{Felder2017} & 0.478 & 1.2$\cdot$10$^5$ & stepped spillway & CP  & 500  & 45  & 5.17 \\
\cite{Felder2019} & 1.086  & 1.1$\cdot$10$^6$ & tunnel chute & CP  & 500  & 45  & 5.06 \\
\cite{Felder2019} & 1.086 & 1.1$\cdot$10$^6$ & tunnel chute & FO  & 500  & 45  & 5.00 \\
\cite{Montano19} &  0.083 & 8.3$\cdot$10$^4$ & hydraulic jump & CP  & 20  & 300  & 7.9 \\
\bottomrule
\end{tabular}
\end{footnotesize}
\label{tab:experimentalconditions}
\end{table}

\subsection{Stepped spillway (EPFL)}
The stepped spillway chute (Laboratory of Hydraulic Constructions, EPFL) was 0.5 m wide, had a slope of 30 degrees and consisted of equal flat steps with height $h$ = 0.03 m. The dimensionless discharge was $d_c$/$h$ = 9.5 (skimming flow), where $d_c$ is the critical flow depth. In the present study, experimental data at step edge 101 were re-analysed. Further details on the flume and the flow conditions can be found in \cite{Felder2017}.

\subsection{Smooth tunnel chute (ETH Zurich)}
The smooth tunnel chute (Laboratory of Hydraulics, Hydrology and Glaciology, ETH Zurich) had a length of 20.6 m, a width of 0.2 m and a height of \mbox{0.3 m.} The flow in the tunnel chute was controlled with a high-head sluice gate, providing high-velocity air-water flows along the tunnel chute \citep{Felder2019}. For the present re-analysis, flow conditions with an upstream head of $H$ = 15 m and relative gate opening of $h_g/h_{\mathrm{us}}$ = 0.4 were used, where $h_g$ is the opening height of the gate and $h_{\mathrm{us}}$ the flow depth upstream of the sluice gate
\citep{Hohermuth2019}. The re-analysed phase-detection probe data were collected 17.92 m downstream of the sluice gate.

\subsection{Hydraulic jump (UNSW Sydney)}
The open channel flume (Water Research Laboratory, UNSW Sydney) used for the hydraulic jump experiments was 40 m long and 0.6 m wide. An upstream sluice gate with a rounded corner controlled supercritical flows and the longitudinal position of the hydraulic jump was adjusted with a tail gate at the end of the flume \citep{Montano19}. In the present study, conductivity probe data collected at 0.24 m distance from the mean jump toe were re-analysed. The inflow was partially developed (jump toe located 0.5 m downstream of the sluice gate) with inflow depth and inflow Froude number of $d_1$ = 0.03 m and $Fr_1$ = 5.1, respectively. 

\section{Results}
\label{sec:results}

\subsection{Sensitivity analysis of processing parameters}
\label{sec:selection}

The application of the AWCC technique requires the selection of processing parameters, including thresholds for $R_{12,i,\mathrm{max}}$ and SPR$_{i}$ and the number of particles $N_p$. Detailed sensitivity analyses were conducted for all data sets, providing consistent parameters irrespective of the air-water flow phenomenon. In this section, the results of the sensitivity analysis are presented for the ETH tunnel chute.

\subsubsection{Filtering criteria for $R_{12,i,\mathrm{max}}$ and SPR$_i$}
\label{sec:filtering}
Erroneous velocities can be eliminated using a cross-correlation based filtering approach with filtering thresholds $R_{12,\mathrm{thresh}}$ and SPR$_{\mathrm{thresh}}$. These thresholds imply that velocity information is valid if $R_{12,i,\mathrm{max}} > R_{12,\mathrm{thresh}}$ and/or if SPR$_i $ $<$ SPR$_{\mathrm{thresh}}$ \citep{Matos2002,Keane1990,Kramer19AWCC}. \cite{Matos2002} and \cite{Andre2005} used 0.5 $\leq$ $R_{12,\mathrm{thresh}}$ $\leq$  0.7 for phase-detection probe measurements and \cite{Keane1990} and \cite{Hain2007} recommended 0.5 $\leq$ SPR$_{\mathrm{thresh}}$ $\leq$ 0.8 for particle image velocimetry (PIV). 

\begin{figure}[h!]
\centering
\includegraphics[width=0.95\textwidth]{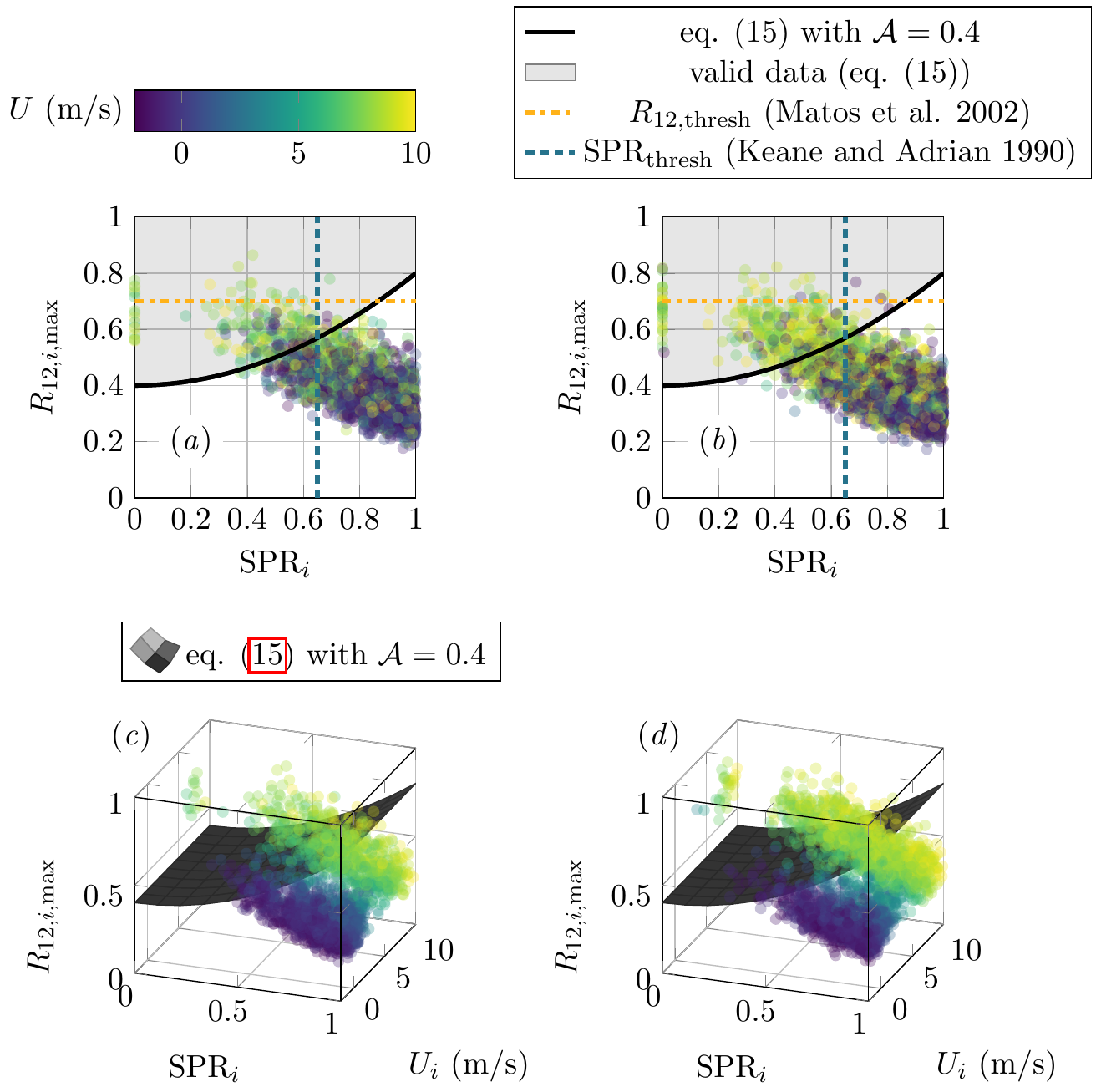}
\caption{Instantaneous $U_i$, $R_{12,i,\mathrm{max}}$ and SPR$_i$ for two measurement locations next to the bottom of the tunnel chute, processed with $N_p$ = 6 (\textit{a}) $R_{12,i,\mathrm{max}}$ versus SPR$_i$ at $y = 4.5$ mm \protect (\textit{b}) $R_{12,i,\mathrm{max}}$ versus SPR$_i$ at $y = 10.5$ mm \protect (\textit{c}) $R_{12,i,\mathrm{max}}$ versus SPR$_i$ versus $U_i$ at $y = 4.5$ mm \protect (\textit{d}) $R_{12,i,\mathrm{max}}$ versus SPR$_i$ versus $U_i$ at $y = 10.5$ mm.} 
\label{SPRvsR12}
\end{figure}

Figure \ref{SPRvsR12} shows two- and three-dimensional representations of pseudo-instantaneous $U_i$, $R_{12,i,\mathrm{max}}$ and SPR$_i$ for two measurement locations next to the bed of the tunnel chute (processed with $N_p$ = 6), including thresholds $R_{12,\mathrm{thresh}}$ = 0.7 \citep{Matos2002} and SPR$_{\mathrm{thresh}}$ = 0.65 \citep{Keane1990}. The magnitude of the instantaneous velocity is plotted using a color scale and dark blue dots represent non-physical near-zero velocities (see section \ref{sec:Np}). It was observed that the secondary peak ratio (SPR$_i$) was linear negatively correlated with the cross-correlation coefficient $R_{12,i,\mathrm{max}}$ (figure \ref{SPRvsR12}). This implies that a velocity estimation with low $R_{12,i,\mathrm{max}}$ has a high probability of also showing a high SPR$_i$ value. Therefore, the use of an empirical criteria $R_{12,i,\mathrm{max}} >  (a \cdot \mathrm{SPR}_i^2) + b$ is proposed, aiming to eliminate erroneous velocity estimations and to   maximise the amount of accepted data through a combination of $R_{12,i,\mathrm{max}}$ and SPR$_i$ (figures \ref{SPRvsR12}\textit{a,b)}. This quadratic inequality can be simplified by setting  $a = b = \mathcal{A}$:

\begin{equation}
\label{eq:filter}
R_{12,i,\mathrm{max}} / \left(\mathrm{SPR}_i^2 + 1 \right) > \mathcal{A} 
\end{equation}
where the parameter $\mathcal{A}$  was chosen as $ \mathcal{A}= 0.4$ (figure \ref{SPRvsR12}). 
This approach implies minimum cross-correlation coefficients of $R_{12,i,\mathrm{max}} > 0.4$ for \mbox{SPR$_i$ = 0} and $R_{12,i,\mathrm{max}} > 0.8$ for SPR$_i$ = 1 (figures \ref{SPRvsR12}\textit{a,b}). While the lowest $R_{12,i,\mathrm{max}}$ and the highest SPR$_i$ values were -- standalone --  less restrictive than in previous studies \citep{Keane1990,Matos2002}, the combination of both parameters provided a robust filtering approach, which was suitable for all investigated flow situations and for both, FO and CP probes. Note that $\mathcal{A}$ in eq. (\ref{eq:filter}) was selected for flow locations next to the solid boundary, representing the most challenging locations due to high turbulence levels, low void fractions and relatively low data yield. 

\subsubsection{Number of particles $N_p$}
\label{sec:Np}
The duration of the adaptive windows is linked to the number of particles in the windows for the leading and trailing tips (figure \ref{fig:probe}\textit{b}). A large window duration can lead to a smoothing/averaging of the velocity fluctuations and therefore to an underestimation of the turbulence levels \citep{Kramer19AWCC}. A more accurate estimation of the pseudo-instantaneous interfacial velocity is therefore expected with smaller window sizes, hence smaller $N_p$. However, a smaller number of particles $N_p$ may lead to non-physical velocity estimations.

\begin{figure}[h!]
\centering
\includegraphics[width=0.95\textwidth]{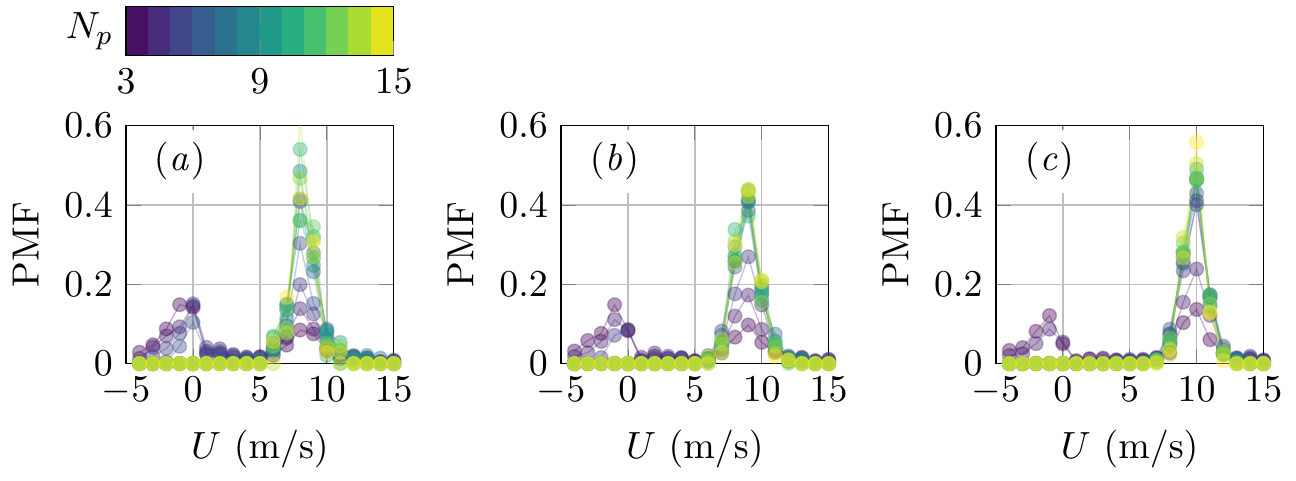}
\caption{Unimodal and (non-physical) bimodal velocity distributions for different number of particles $N_p$ in a high-speed air-water flow down a tunnel chute close to bottom level, processed with $\mathcal{A} = 0.4$ (\textit{a}) $y = 4.5$ mm; $C$ = 0.028; $F$ = 362 s 1/s \protect (\textit{b}) $y = 10.5$ mm; $C$ = 0.045;  $F$ = 493 s$^{-1}$ \protect (\textit{c}) $y = 20.5$ mm; $C$ = 0.066;  $F$ = 638 s$^{-1}$.}
\label{fig:PMF}
\end{figure}

In turbulent air-water flows, three-dimensional motion increases the probability that particles are detected only by one tip. In very small windows, random solitary interfaces weight considerably, sometimes resulting in velocity PMFs around zero, which is a non-physical estimation (figures \ref{SPRvsR12}, \ref{fig:PMF}). This behaviour was observed in regions with high turbulence intensities, e.g. close to a solid boundary. The window duration, and therefore $N_p$, must be large enough to ensure that a single interfacial event cannot impair the velocity estimation. Using eq. (\ref{eq:filter}) with $\mathcal{A} = 0.4$, a number of particles 5 $\leq$ $N_p$ $\leq$ 15 was found appropriate to eliminate non-physical velocities close to zero (figure \ref{fig:PMF}).  

\subsection{Boundary layer flows: ETH tunnel chute and EPFL stepped spillway}
The phase-detection probe datasets of the smooth tunnel chute (CP and FO) and the  stepped spillway (CP) were processed with the AWCC technique following section \ref{sec:signalprocessing}. The number of particles was varied between 5 $\leq N_p \leq$ 15 and $\mathcal{A}=0.4$ was used for cross-correlation based filtering (eq. (\ref{eq:filter})). Similar to \cite{Straub58} and \cite{Zhang2017}, vertical coordinates were normalised with $Y_{50}$, i.e.  the elevation where $C = 0.5$.

\subsubsection{Mean velocity and velocity fluctuations}
\begin{figure}[h!]
\centering
\includegraphics[width=0.95\textwidth]{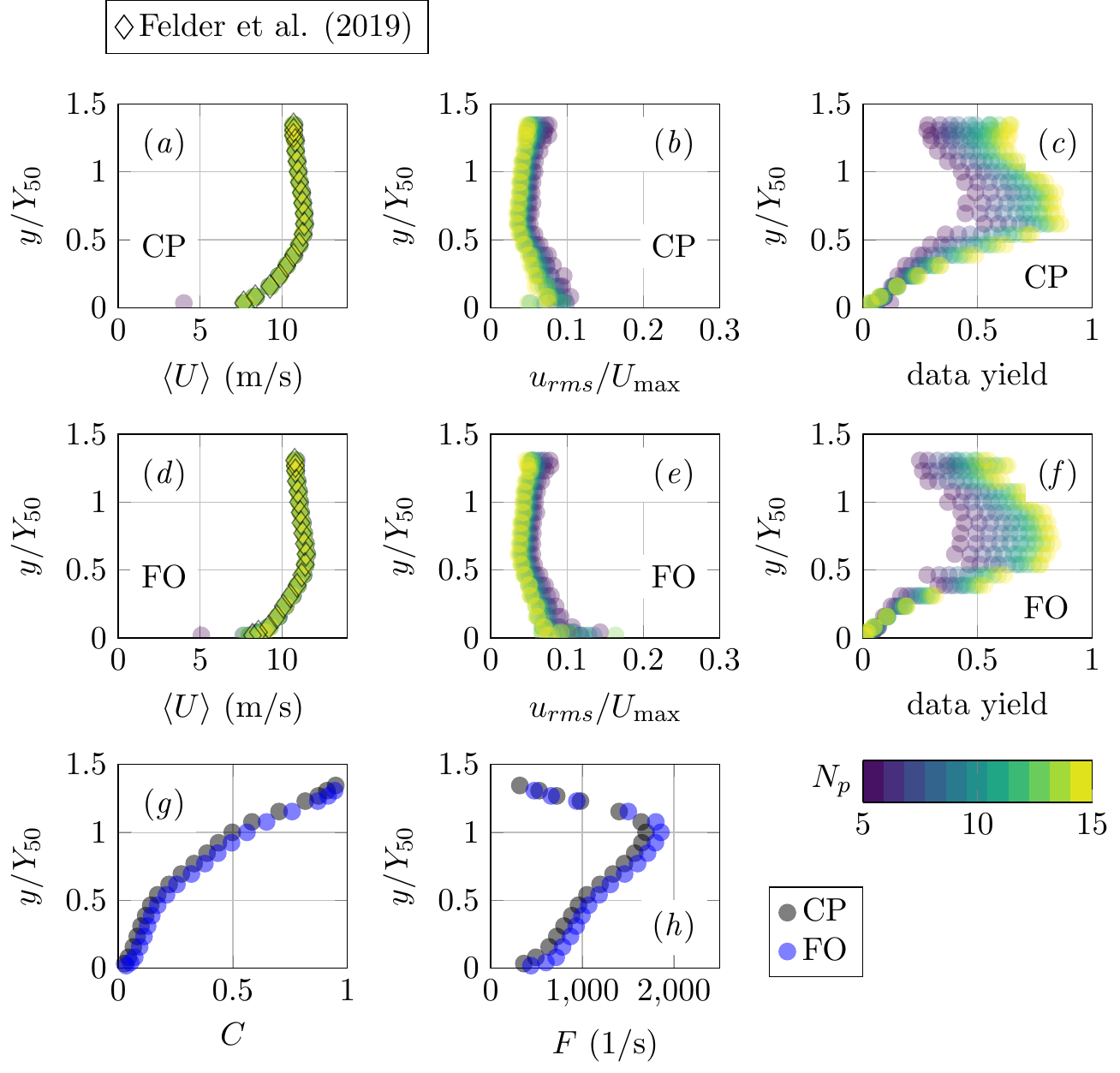}
\caption{Tunnel chute: application of the AWCC technique with different $N_p$ and $\mathcal{A}=0.4$; conductivity probe (CP) versus fiber optical (FO) probe (\textit{a}) CP: mean velocities (\textit{b}) CP: turbulent fluctuations (\textit{c}) CP: data yield (\textit{d}) FO: mean velocities (\textit{e}) FO: turbulent fluctuations (\textit{f}) FO: data yield (\textit{g}) void fraction (\textit{h}) particle count rate.}
\label{fig:ETH1}
\end{figure}

\begin{figure}[h!]
\centering
\includegraphics[width=0.95\textwidth]{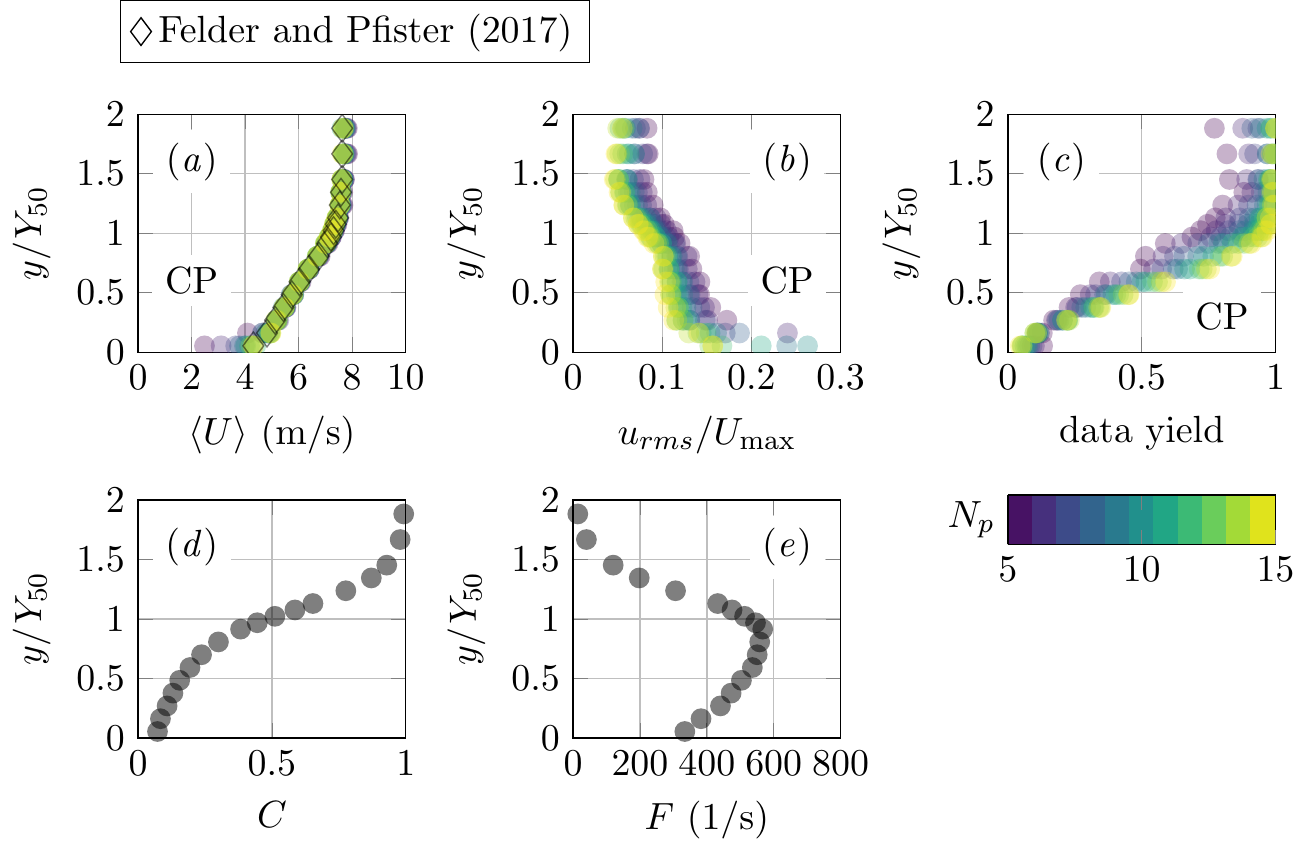}
\caption{Stepped spillway: application of the AWCC technique with different $N_p$ and $\mathcal{A}=0.4$; measurements undertaken with conductivity probe (CP) (\textit{a}) mean velocities (\textit{b}) turbulent fluctuations (\textit{c}) data yield (\textit{d}) void fraction (\textit{e}) particle count rate.}
\label{fig:EPFL1}
\end{figure}

Mean velocities $\langle U \rangle$ obtained with the AWCC were in close agreement with those of the conventional cross-correlation technique for the re-analysed datasets (figures \ref{fig:ETH1}\textit{a},\textit{d} and \ref{fig:EPFL1}\textit{a}). 

As shown in \cite{Kramer19AWCC}, the estimated mean velocities  were independent of the number of particles $N_p$. The velocity profile in the tunnel chute showed a velocity-dip, typical for small aspect ratios \citep{Hohermuth2019}, whereas the profile on the stepped spillway followed closely a power law \citep{Chanson13}. The scatter in $\langle U \rangle$ close to the bed of both chutes was due to low data rates and is anticipated to disappear with longer sampling durations. For the tunnel chute, the mean velocity profiles measured with the conductivity probe (CP) and the fibre-optical probe (FO) were similar (figures \ref{fig:ETH1}\textit{a,d}). However, the FO probe resulted in slightly larger velocities close to the chute bottom, which is expected to be related to the smaller probe tips \citep{Vejrazka2010}.

Turbulent fluctuations were evaluated on the basis of eq. (\ref{eq:rms}), normalised with the maximum cross-sectional velocity $U_{\mathrm{max}}$ (figures \ref{fig:ETH1}\textit{b},\textit{e} and \ref{fig:EPFL1}\textit{b}). High fluctuations were found next to the bed of the tunnel chute and close to the  pseudo-bottom of the stepped spillway. The streamwise normal Reynolds stresses were similar to boundary layer flows \citep{Pope20}, albeit with slightly larger turbulence levels in air-water flows. The velocity fluctuations were higher on the stepped spillway compared to the tunnel chute due to the macro-roughness of the steps, showing a small bump at lower depths. 

The scatter for 5 $\leq$ $N_p$ $\leq$ 6 was due to the bimodal velocity distributions observed for small $N_p$ (see figure \ref{fig:PMF}), whereas the decrease in turbulence levels for $N_p$ $\geq$ 7 was due to averaging effects with increasing window duration (\ref{app:extrapolation}). There was no significant difference in measured turbulence intensities for CP and FO probes, confirming that the AWCC is applicable to different types of phase-detection probes.

The data yield was approximately 0.05 $<$ data yield $<$ 0.95 (figures \ref{fig:ETH1}\textit{c,f} and \ref{fig:EPFL1}\textit{c}) and increased with distance from the bottom, which was directly related to the distribution of void fractions and particle count rates (figures \ref{fig:ETH1}\textit{g,h} and \ref{fig:EPFL1}\textit{d,e}), combined with greater probabilities of single-tip impacts in flow regions with high turbulence intensities. 

\subsubsection{Time-resolving turbulent properties: a word of caution}
\begin{figure}[h!]
\centering
\includegraphics[width=0.95\textwidth]{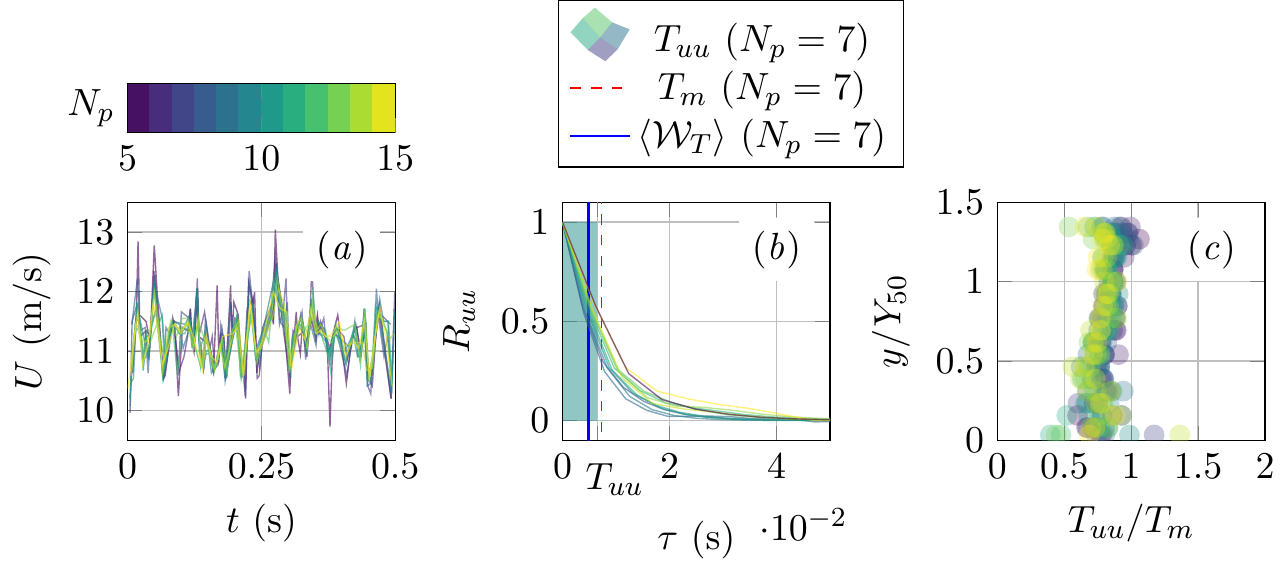}
\caption{Time-resolving turbulent properties (tunnel chute) for different $N_p$, recorded with a conductivity probe (CP) (\textit{a}) pseudo-instantaneous velocity time series at $y =$ 100.5 mm (only 0.5 s are shown for clarity) (\textit{b}) ACFs, integral time scale ($T_{uu}$), measurement time scale  ($T_{m}$) and mean window duration ($\langle \mathcal{W}_T  \rangle$) at $y =$ 100.5 mm  (\textit{c}) data density ($T_{uu}/T_m$) across the air-water column.}
\label{fig:spectra_tunnel}
\end{figure}

The calculation of pseudo-instantaneous interfacial velocities with the AWCC technique may allow to uncover further turbulence properties. Figure  \ref{fig:spectra_tunnel}\textit{a} shows pseudo-instantaneous velocity time series for different number of particles, recorded at a distance of $y$ = 100.5 mm from the bed of the tunnel chute. As discussed in section \ref{sec:Np}, increasing $N_p$ resulted in a smoothing of the velocity time series. 

Auto-correlation functions (ACF) for different numbers of particles were calculated using NN-resampling  and ARMA model fitting (section \ref{sec:timeseries}) (figure \ref{fig:spectra_tunnel}\textit{b}). The ACFs demonstrated a subtle trend of increasing auto-correlation coefficient ($R_{uu}$) with increasing $N_p$, which might be due to the stronger influence of resampling at lower data rates or larger $N_p$, respectively. The integral velocity time scale ($T_{uu}$) represents a measure of the \textit{memory} of the flow \citep{Kundu2016} and was calculated by integrating the ACF to the first zero-crossing:
\begin{equation}
T_{uu} = \int_{\tau = 0}^{\tau(R_{uu}=0)} R_{uu}(\tau) d\tau
\label{eq:integraltimescale}
\end{equation}
with $\tau$ the lag time. As shown in figure \ref{fig:spectra_tunnel}\textit{b}, the integral time scale $T_{uu}$ is equal to the area under a rectangle of unity height. The measurement time scale $T_m = 1/f_m$ is defined as the reciprocal of the mean data rate 
(figure \ref{fig:spectra_tunnel}\textit{b}). To avoid velocity bias and aliasing, a data density (= ratio of the integral time scale to the measurement time scale) in the order of $T_{uu}/T_m > 5$ \citep{Winter1991} or $T_{uu}/T_m > 10$ \citep{Damaschke2018} is typically recommended. 

In this study, it was observed that $T_{uu}/T_m$ was slightly below unity across the full air-water flow column of the tunnel chute (figure \ref{fig:spectra_tunnel}\textit{c}). This finding implies that the AWCC data density did not allow for a reliable quantification of time-resolving turbulent properties. For $N_p = 7$, even a 100 \% data yield would be insufficient for an unbiased estimation of $T_{uu}$, i.e. $T_{uu}/\langle W_T \rangle < 10$ (figure \ref{fig:spectra_tunnel}\textit{b}). Moving towards single event detection techniques or improving the design of the probes may help to overcome these limitations in the future. 

\subsection{Shear layer flow: the UNSW hydraulic jump}
\begin{figure}[h!]
\centering
\includegraphics[width=0.95\textwidth]{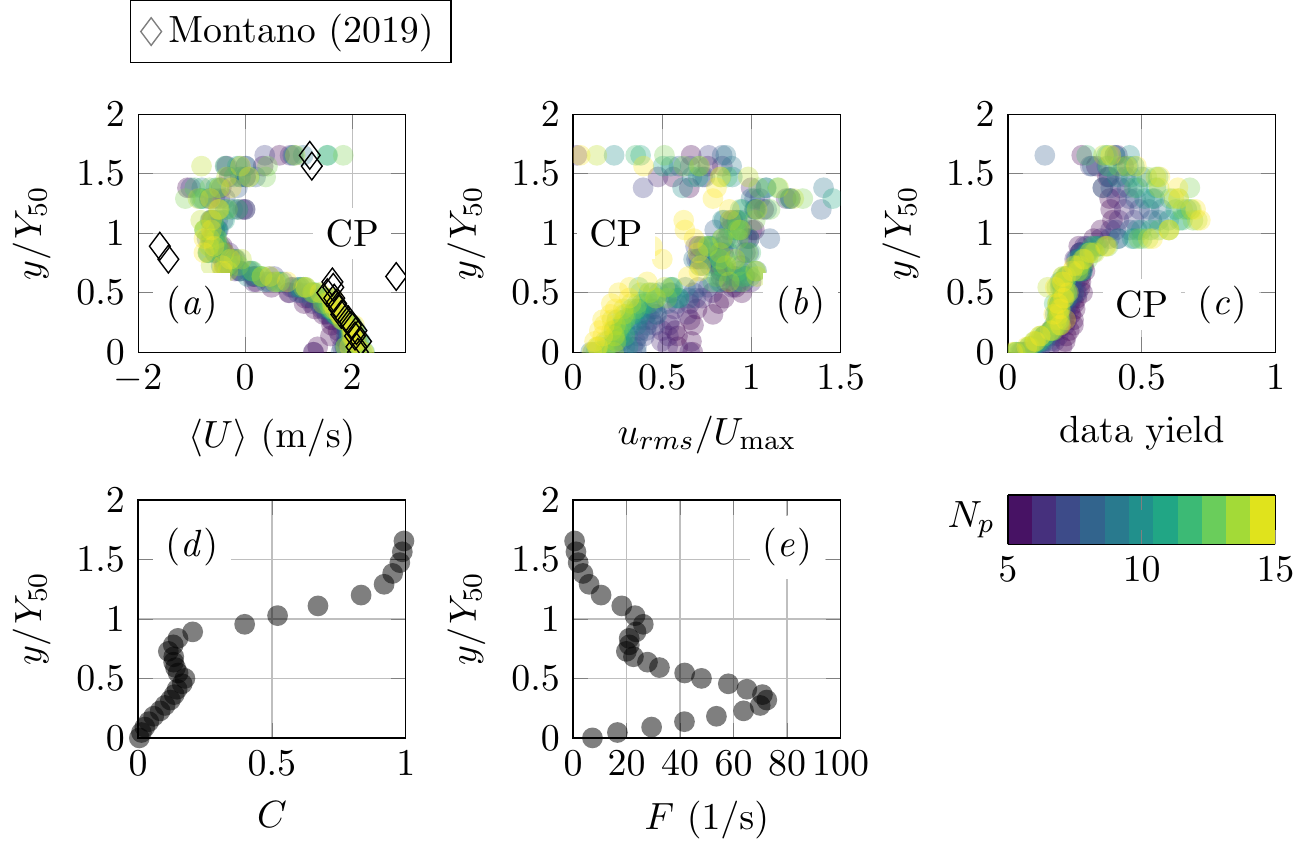}
\caption{Hydraulic jump: application of the AWCC technique with different $N_p$ and $\mathcal{A}=0.4$; measurements undertaken with a conductivity probe (CP) (\textit{a}) mean velocities (\textit{b}) turbulent fluctuations (\textit{c}) data yield (\textit{d}) void fraction (\textit{e}) particle count rate.}
\label{fig:WRL1}
\end{figure}

The hydraulic jump is a well-known hydraulic phenomena that constitutes a transition from supercritical to subcritical flow conditions.
The measured void fraction and particle count rate  distributions showed typical profiles with two peaks (figures \ref{fig:WRL1}\textit{d},\textit{e}), one in the shear layer and one in the recirculation region \citep{Murzyn05, Wang15,Montano19}. 

Due to the complex nature of fully aerated hydraulic jumps, the measurement of interfacial velocities is challenging and limitations exist in the upper flow region, which is due to three-dimensional velocity contributions. Herein, the AWCC was applied to provide mean interfacial velocities and turbulence levels. Next to the channel bed, the interfacial velocity distribution was similar to a wall-jet \citep{Rajaratnam65}. Largest velocities were observed for the impinging jet, followed by a shear layer and on average negative velocities in the recirculation region (figure \ref{fig:WRL1}\textit{a}). A continuous velocity profile was attained irrespective of $N_p$, which resembled the results presented by \cite{Zhang14} and \cite{Kramer2019HJ}. 

The turbulence levels indicated significant turbulent interfacial interactions with largest levels within the recirculation region (figure \ref{fig:WRL1}\textit{b}). It must be noted that a misalignment of the probe tips with flow streamlines, especially in the recirculation region, and high turbulence levels may have led to an increase of the measurement uncertainty (see also table \ref{tab:weighting}). Despite a data yield $<$ 0.7 over the whole water column (and below 0.3 for $y<80$ mm), the shape of interfacial velocity and turbulence intensity distributions remained continuous for a sampling duration of \mbox{$T$ = 300 s.}

\section{Discussion}
\label{sec:discussion}
\subsection{Comparison to conventional processing techniques}

The AWCC technique \citep{Kramer19AWCC} with present modifications offers an alternative to conventional methods described in \mbox{section \ref{sec:conventional}.} With the conventional method, interfacial mean velocities are typically evaluated from cross-correlation analysis of the complete raw signal \citep{Toombes2002} or from an ensemble-averaged analysis of sub-segments with a sub-sample duration between 3 to 5 s  \citep{Felder2015}. Conventional turbulence level estimations are based on eq. (\ref{eg:Tuold}), proposed by \cite{Chanson2002a}.

\begin{figure}[h!]
\centering
\includegraphics[width=0.95\textwidth]{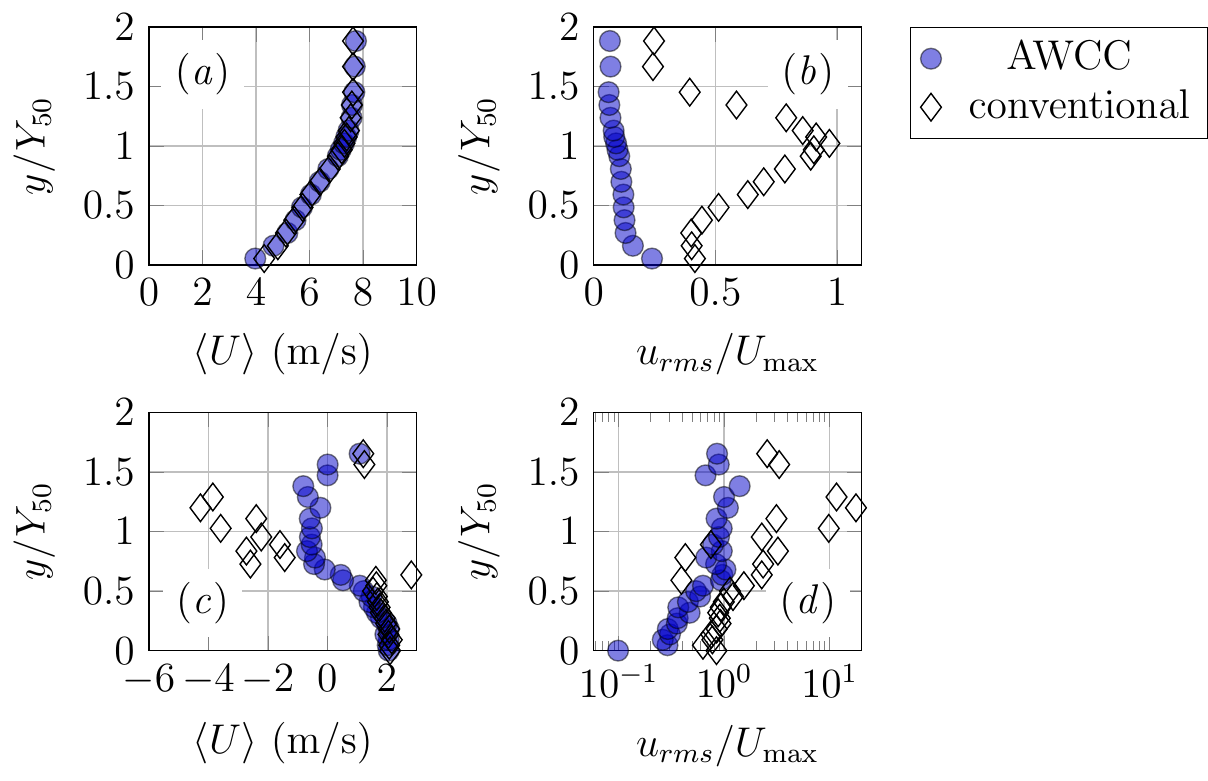}
\caption{Comparison of the AWCC technique ($N_p$ = 10 and $\mathcal{A}=0.4$) with conventional signal processing after \cite{Chanson2002a}; measurements undertaken with conductivity probes (CP) (\textit{a}) stepped spillway: mean velocity profiles \protect (\textit{b})  stepped spillway: streamwise turbulent fluctuations \protect (\textit{c}) hydraulic jump: mean velocity profiles  (\textit{d}) hydraulic jump: streamwise turbulent fluctuations.}
\label{fig:comparison}
\end{figure}

In high-velocity air-water flows down stepped and smooth chutes with a clearly defined main flow direction, the mean flow velocities estimated with the AWCC technique and the conventional approach were in close agreement (figure \ref{fig:comparison}\textit{a}). However, the conventional mean velocity estimation is based on the analysis of a \textit{single} ensemble-averaged cross-correlation function, which does not allow for the detection of fluctuating signs and near-null velocities \citep[as discussed by][]{Wang15}. In shear layer flows, the conventional processing therefore resulted in an non-physical overestimation of the recirculation velocity  (figure \ref{fig:comparison}\textit{c}). The AWCC allowed the consistent characterisation of the jump's shear layer and the results were similar to observations of a jump with $Fr_1$ = 5.43 \citep{Zhang14}. Note that the AWCC results were further validated via image-based velocimetry from a sidewall perspective by \cite{Kramer2019HJ}.

Streamwise turbulence levels computed with the conventional approach were significantly larger than the AWCC results (figure \ref{fig:comparison}\textit{b}). While the AWCC data followed a profile similar to a single-phase flow with maximum turbulence intensities close to the pseudo-bottom \citep[as demonstrated in][]{Amador2006}, the method of \cite{Chanson2002a} resulted in a local maximum in the intermediate flow region at \mbox{$y$ $\approx$ 100 mm} (figure \ref{fig:comparison}\textit{b}). This overestimation is a result of the shortcomings of eq. (\ref{eg:Tuold}) as discussed in section \ref{sec:conventional}.
Unrealistically high turbulence levels of $u_{rms}/U_{\mathrm{max}}$ $\approx$ 18 (figure \ref{fig:comparison}\textit{d}) were also observed with the conventional method for shear layer flows in the hydraulic jump \citep[see also][]{Zhang14}. In contrast, the AWCC technique provided continuous profiles of much lower turbulence levels.

Based on the observations made herein, the conventional method should only be applied to calculate mean streamwise velocities in air-water flows with mostly unidirectional flow velocity (i.e. spillways and chutes), while conventional turbulence levels must be treated with caution for any type of air-water flow. The AWCC provides a more consistent estimation of near-null velocities and turbulent fluctuations, but is also not completely free of biases; e.g. particle-probe interaction and a variation of the angle of attack are believed to increase the measurement uncertainty \citep{Thorwarth2008,Vejrazka2010}. The use of synthetic signals and the comparison to other, preferably non-intrusive, measurements techniques can help to refine the technique in future works.

\subsection{Overcoming the window-smoothing effect: extrapolation of turbulence levels to $N_p$ = 1}
\begin{figure}[h!]
\centering
\includegraphics[width=0.95\textwidth]{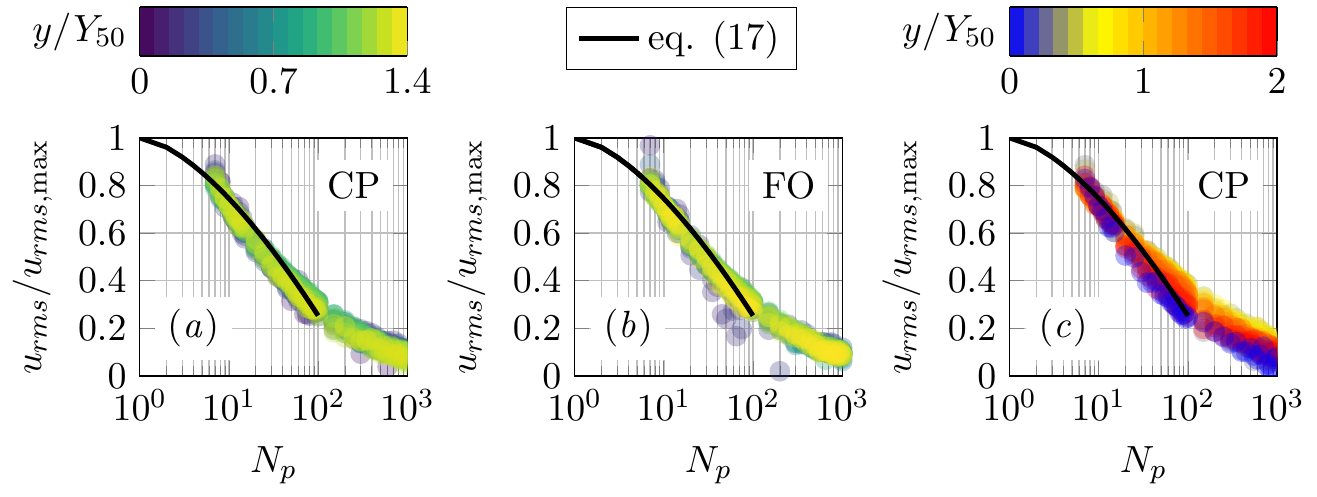}
\caption{Normalised turbulent fluctuations ($u_{rms}/u_{rms,\mathrm{max}}$) versus $N_p$ for the tunnel chute and the stepped spillway, processed with  $\mathcal{A}=0.4$ (\textit{a}) conductivity probe (CP) measuremements of the tunnel chute (\textit{b}) fiber optical probe (FO) measuremements of the tunnel chute (\textit{c}) conductivity probe (CP) measuremements of the stepped spillway.}
\label{fig:Np}
\end{figure}

To overcome the  practical limitation of processing with a very small number of particles, the convergence of velocity fluctuations for single particles ($N_p$ = 1) was investigated for the chute datasets. 
Figure \ref{fig:Np} depicts measured turbulence intensities versus $N_p$.   
Data points with very low or very high void fractions, e.g. close to the bed of the chutes or in the upper flow region, were excluded from the analysis due to potentially non-physical velocities and insufficient data rates. Estimated fluctuations decreased with increasing $N_p$ and followed an expression derived using stochastic signals (\ref{app:extrapolation}):
\begin{equation}
u_{rms}/u_{rms,\mathrm{max}} = 1 - \mathcal{K}_1  \ln (N_p)^{\mathcal{K}_2} \quad \quad \{ N_p \in \mathbb{Z}^{+}: 1 \leq N_p \leq 100 \}
\label{eq:fit}
\end{equation}
where $u_{rms,\mathrm{max}}$ is a parameter that represents the extrapolated root mean square of velocity fluctuations for $N_p$ = 1 at each depth $y$. The equation was valid up to $N_p$ $\approx$ 100 and the constants took the values  $\mathcal{K}_1 = 0.07$ and $\mathcal{K}_2 = 1.55$ for the available data (figure \ref{fig:Np}).

Independent of the type of phase-detection probe, eq. (\ref{eq:fit}) showed good agreement for different types of air-water flows. The obtained values imply that velocity fluctuations for $N_p$ = 10 are roughly 70 \% of those ideally expected for $N_p$ = 1, and 30 \% for $N_p$ = 100. It is stressed that the constants $\mathcal{K}$ are a function of the particle count rate $F$, the turbulence level $Tu$ and the integral time scale $T_{uu}$ of the flow. Therefore, it is recommended to perform a calibration before extrapolating turbulence levels with eq. (\ref{eq:fit}).

\subsection{Best practices for the AWCC technique}
While the AWCC technique cannot solve all limitations of phase-detection probes, it helps alleviating shortcomings of conventional analysis methods. The sensitivity analysis of the AWCC technique for three different flow types revealed that the optimum processing parameters are in the same range for all tested flow conditions and probe types (side-by-side tip design). Best practice recommendations for the AWCC technique are as follows:

\begin{enumerate}
\item Select an appropriate cross-correlation based filtering approach, starting with eq. (\ref{eq:filter}) using $\mathcal{A} = 0.4$.

\item Select the number of particles $N_p$ within a recommended range between 5 $\leq N_p \leq 15$. It is advised to perform a sensitivity analysis similar to figure \ref{fig:PMF}, with the aim to keep $N_p$ small while avoiding non-physical velocity information (i.e. bimodal PMF).

\item Correct the velocity bias using a suitable weighting scheme (equations (\ref{eq:meanvelocity}) and (\ref{eq:rms})). Window duration weighting and interarrival time weighting showed a similar performance, whereas inverse absolute velocity weighting led to unreasonable mean velocities in free-shear regions characterised by near-zero velocities. 

\item When the focus is on turbulence quantities, eq. (\ref{eq:fit}) can be used to extrapolate turbulence intensities to $N_p$ = 1. Because the constants $\mathcal{K}$ may be a function of sampling parameters and flow properties, a calibration is necessary. 

\item  A sufficient number of valid pseudo-instantaneous velocity data must be recorded to reduce the uncertainty in the estimation of mean and turbulent properties (\ref{app:convergence}). A low data yield directly implies the need for longer sampling durations. In such cases, it is recommended to increase the sampling duration until the moments of the velocity samples converge.
\end{enumerate}

\section{Conclusion}
This study presented a detailed sensitivity analysis of the adaptive window cross-correlation (AWCC) technique for processing dual-tip phase-detection probe signals in highly-aerated flows. The re-analysed experimental datasets covered three common air-water flow phenomena, including a smooth tunnel chute, a stepped spillway and a hydraulic jump.

It was observed that a cross-correlation based filtering approach performed well for all flow conditions, regardless of the deployed dual-tip phase-detection probe (fiber optical and conductivity probe) with side-by-side tip design. Mean velocities were independent of the number of particles ($N_p$) per window, but estimated velocity fluctuations decreased with increasing $N_p$. Stochastic signals were used to derive an empirical expression that allows the extrapolation of turbulence levels to $N_p=1$, thereby solving the practical limitation that recorded signals cannot be processed with very small numbers of particles.

A comparison with conventional signal processing confirmed that these methods were not able to measure near-null velocities, as those in the shear layer of hydraulic jumps. As mentioned in \cite{Zhang14}, the computation of turbulence levels from standard deviations of auto- and cross-correlation functions (of the raw voltage signals) seemed to significantly overestimate the turbulence intensity and published data must be treated with caution.

The AWCC technique allows a reliable analysis of mean interfacial velocities and turbulent fluctuations in highly aerated flows. Its ability to uncover time-resolving turbulence statistics is currently limited by the data density. Future developments in terms of signal processing and optimized design of phase-detection probes may help to overcome these limitations. Best practices have been proposed to provide a clear guidance for future studies using the AWCC technique.

\section*{Acknowledgements} 
The authors would like to thank Dr Laura Montano (WRL, UNSW Sydney) for providing the hydraulic jump data set and Rob Jenkins (WRL, UNSW Sydney) for manufacturing the WRL conductivity probes. We also would like to thank the anonymous reviewers for their valuable comments, which helped to improve the manuscript.

\section*{Notation} \label{sec:notation}
\vspace{-0.5cm}
\begin{tabbing}
\hspace*{0cm}\=\hspace*{2.2cm}\=\hspace*{10.5cm}\=\kill\\
\>$\mathcal{A}$  \> parameter for cross-correlation based filtering \> (-)\\
\>$a$  \> constant  \> (-)\\
\>$b$  \> constant  \> (-)\\
\> $c(t)$  \> instantaneous void fraction (= binarized signal) \> (-)\\
\>$C$  \> time-averaged void fraction \> (-)\\
\> $d_c$  \> critical flow depth (stepped spillway) \> (m)\\
\> $d_1$  \> inflow depth (hydraulic jump) \> (m)\\
\> $e_{\langle U \rangle}$  \> relative error; $e_{\langle U \rangle} = \left(\langle U \rangle - \langle U \rangle_{\mathrm{sim}}\right)/\langle U \rangle_{\mathrm{sim}}$ \> (-)\\
\> $e_{Tu}$  \> relative error; $e_{Tu} = \left(Tu - Tu_{\mathrm{sim}}\right)/Tu_{\mathrm{sim}}$ \> (-)\\
\>$f$  \> sampling rate \> (s$^{-1}$)\\
\>$f_{m}$  \> data rate (valid windows per time) \> (s$^{-1}$)\\
\>$F$  \> time-averaged particle count rate \> (s$^{-1}$)\\
\>$Fr_1$  \> inflow Froude number (hydraulic jump) \> (-)\\
\>$g$  \> gravitational acceleration \> (m s$^{-2}$)\\
\>$h_{\mathrm{us}}$  \> flow depth upstream of sluice gate (tunnel chute)\> (m)\\
\>$h$  \> step height (stepped spillway)\> (m)\\
\>$h_g$  \> sluice gate opening (tunnel chute)\> (m)\\
\>$H$  \> upstream head (tunnel chute) \> (m)\\
\>$\mathcal{K}$  \> constants for extrapolating $rms$ velocity fluctuations \> (-)\\
\>$n_{\mathcal{W}}$  \> number of windows \> (-)\\
\>$m_a$  \> mode representing the air phase \> (V)\\
\>$m_w$  \> mode representing the water phase \> (V)\\
\>$N$  \> number of detected particles  \> (-)\\
\>$N_e$  \>  ensemble size\> (-)\\
\>$N_p$  \>  selected number of particles to determine window sizes \> (-)\\
\>$q$  \> specific water discharge \> (m$^2$ s$^{-1}$)\\
\>$Re$  \> Reynolds number \> (-)\\
\>$R_{uu}$  \> auto-correlation coefficient \> (-)\\
\>$R_{12}$  \> cross-correlation coefficient \> (-)\\
\>$R_{12,\mathrm{thresh}}$  \> threshold for the cross-correlation coefficient \> (-)\\
\>$S(\bm{x},t)$  \> raw phase-detection probe signal \> (V)\\
\>SPR  \> secondary peak ratio  \> (-)\\
\>SPR$_\mathrm{thresh}$  \> threshold for the secondary peak ratio \> (-)\\
\>$T$  \> sampling duration \> (s)\\
\>$T_{m}$  \> measurement time scale, $T_{m}=1/f_{m}$ \> (s)\\
\>$T_{uu}$  \> integral velocity time scale \> (s)\\
\>$\mathcal{T}_i$  \> interfacial travel time  \> (s)\\
\>$Tu$  \> turbulence intensity, defined as $Tu = u_{rms}/\langle U \rangle$ \> (-)\\
\>$t$  \> time \> (s)\\
\>$t_{ch,a}$  \> air chord time \> (s)\\
\>$t_{ch,w}$  \> water chord time \> (s)\\
\>$t_{s,i}$  \> start time of window $i$ \> (s)\\
\>$t_{e,i}$  \> end time of window $i$ \> (s)\\
\>$U$  \> instantaneous streamwise velocity \> (m s$^{-1}$)\\  
\>$\langle U \rangle $  \> time-averaged streamwise velocity \> (m s$^{-1}$)\\ 
\>$ U_{\mathrm{max}} $  \> maximum time-averaged cross-sectional velocity \> (m s$^{-1}$)\\     
\>$u$  \> fluctuating part of the streamwise velocity \> (m s$^{-1}$)\\
\>$\mathcal{W}_{T}$  \> window duration  \> (s) \\ 
\>$w$  \> weighting factor for velocity bias correction \> (-) \\  
\>$x$  \> streamwise coordinate \> (m)\\   
\>$\bm{x}$  \> coordinate vector \> \\ \>$Y_{50}$  \> vertical elevation where $C = 0.5$ \> (m)\\ 
\>$y$  \> vertical coordinate \> (m)\\ 
               
\end{tabbing}
Greek letters
\vspace{-0.5cm}
\begin{tabbing}
\hspace*{0cm}\=\hspace*{2.2cm}\=\hspace*{10.5cm}\=\kill\\
\>$\Delta x $  \> streamwise tip separation \> (m)\\
\>$\nu $  \> kinematic viscosity of water\> (m$^2$/s)\\
\>$\sigma $  \> standard deviation \> (s)\\
\>$\tau $  \> time lag of the auto- or cross-correlation function \> (s)\\
\>$\phi $  \> diameter \> (mm)\\
\end{tabbing}
Abbreviations, operators and subscripts
\vspace{-0.5cm}
\begin{tabbing}
\hspace*{0cm}\=\hspace*{2.2cm}\=\hspace*{10.5cm}\=\kill\\
\>$\langle ... \rangle $  \> time-average over measurement time $T$\\
\>$\lfloor \cdot \rfloor$  \> floor function \\
\>1, 2  \> leading and trailing tip\\
\>ACF  \> autocorrelation function \\
\>ARMA  \> autoregressive moving-average model \\
\>AWCC  \> adaptive window cross correlation \\
\>CP  \> phase-detection conductivity probe \\
\>FO  \> phase-detection fiber optical probe \\
\>$i$  \> running variable  \> \\
\>$j$  \> running variable  \> \\
\>LDA  \> laser Doppler anemometer  \> \\
\>max  \> maximum  \> \\
\>NN  \> nearest-neighbor resampling  \> \\
\>PIV  \> particle image velocimetry  \> \\
\>PMF  \> probability mass function  \> \\
\>PSD  \> power spectral density function  \> \\
\>$rms$  \> square root of mean velocity fluctuations\> \\
\>ROC  \> robust outlier cutoff \\
\>sim  \> simulation \\
\end{tabbing}

\bibliography{Bib}

\appendix
\section{Convergence analysis - mean and turbulent properties}
\label{app:convergence}
\begin{figure}[h!]
\centering
\includegraphics[width=0.95\textwidth]{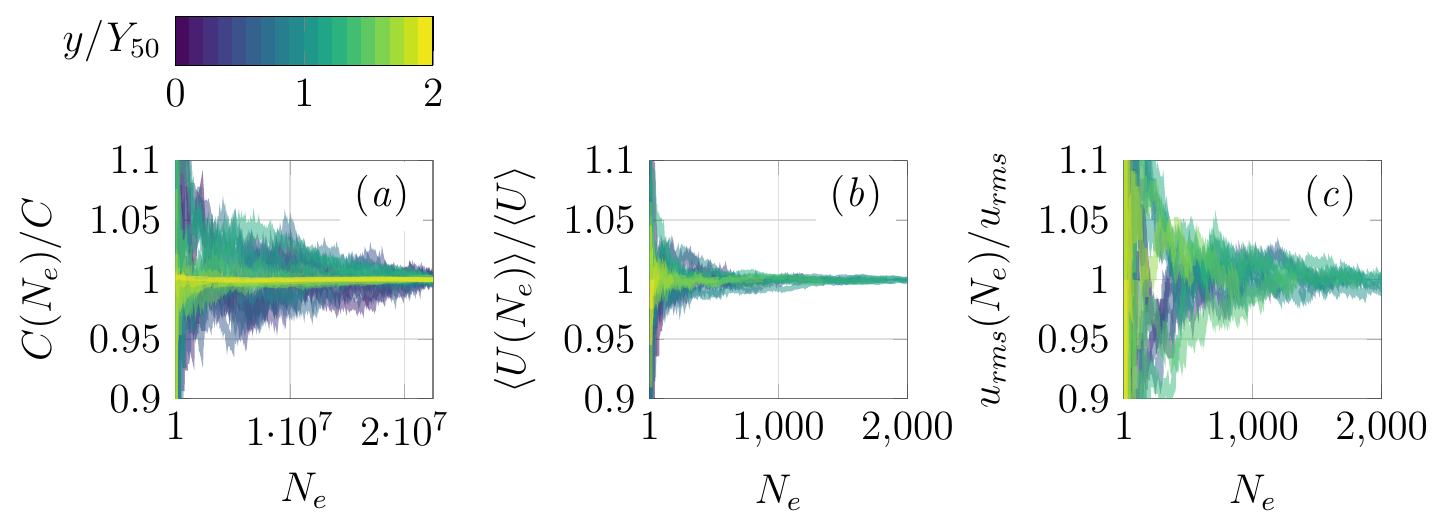}
\caption{Convergence analysis for mean and turbulent properties of the stepped spillway (\textit{a}) mean void fraction (\textit{b}) mean streamwise velocities, computed for $N_p=10$ (\textit{c}) square root of mean velocity fluctuations, computed for $N_p=10$.}
\label{fig:convergence}
\end{figure}

The convergence analysis for mean void fraction $C$, mean flow velocity $\langle U \rangle$ and root-mean-square velocity fluctuations $u_{rms}$ was performed using the stepped spillway data. Running averages of $C(N_e)$, $\langle U(N_e) \rangle$, $u_{rms}(N_e)$ were normalised with mean values $C$, $\langle U \rangle$, $u_{rms}$ of the entire ensemble $N_{e,\mathrm{max}}$ (i.e.  $C(N_{e,\mathrm{max}}$) = $C$) where $N_e$ = ensemble size (figure \ref{fig:convergence}). The entire/maximum ensemble size for the computation of the void fraction was relatively large ($N_{e,\mathrm{max}} = f \cdot T = 2.25 \cdot 10^7$, with $f$ the sampling rate and $T$ the sampling duration) and the best convergence was achieved in the upper flow region (figure \ref{fig:convergence}\textit{a}), where turbulence quantities were the lowest. Slower convergence was observed close to chute invert and uncertainties of roughly $ \pm$2 $\%$ remained for the full sampling duration $T= 45$ s.

The maximum ensemble sizes for the computation of mean velocities and velocity fluctuations were determined by the local particle count rate, the number of particles ($N_p$ = 10) and the local data yield ($N_{e,\mathrm{max}}(y) = n_{\mathcal{W}}(y) \cdot \mathrm{data} \, \mathrm{yield}(y)$, with $n_{\mathcal{W}}$ the number of windows). Note that the maximum ensemble sizes (i.e. the maximum values on the abscissae for each curve in figure \ref{fig:convergence}) represent a sampling duration of $T$ = 45 s. Although $n_{\mathcal{W}} \cdot \mathrm{data} \, \mathrm{yield} << f \cdot T$, mean velocities converged relatively fast at around $N_e \approx 1000$ irrespective of the measurement location (figure \ref{fig:convergence}\textit{b}), whereas the convergence of the velocity fluctuations was slower compared to the mean velocities (figure \ref{fig:convergence}\textit{c}). Overall, the largest ensemble size was required next to the chute invert due to the strongest turbulent fluctuations and three-dimensional motions within this flow region. A sampling duration $T > 45$ s is necessary to achieve convergence uncertainties below $\approx \pm 2\%$.

\section{Extrapolation of turbulence levels - stochastic approach}
\label{app:extrapolation}

The extrapolation of velocity fluctuations to single particles ($N_p$ = 1) was derived using stochastic modelling. Turbulent velocity time series were generated using the Langevin equation \citep{Langevin1908}, governing a stochastic process $u^*$ with mean zero and integral time-scale $T_{uu}$ \citep{Pope20}:

\begin{equation}
 u^* \left(
 t + \delta \, t
 \right) = u^* (t) \left(1 - \frac{\delta\, t}{T_{uu}}
 \right) + u^*_{rms}
 \left( \frac{2 \, \, \delta \, t}{T_{uu}}
 \right)^{1/2}
 \xi (t)
\label{eq:stoch}
\end{equation}
for time steps $\delta\, t \ll T_{uu}$, where $\xi (t)$ is a standardized Gaussian random variable. The modelled velocity $u^*$ represents the fluctuating particle velocity $u$ and was superimposed to a time-averaged velocity $\langle U \rangle$. 

\begin{figure}[h!]
\centering
\includegraphics[width=0.95\textwidth]{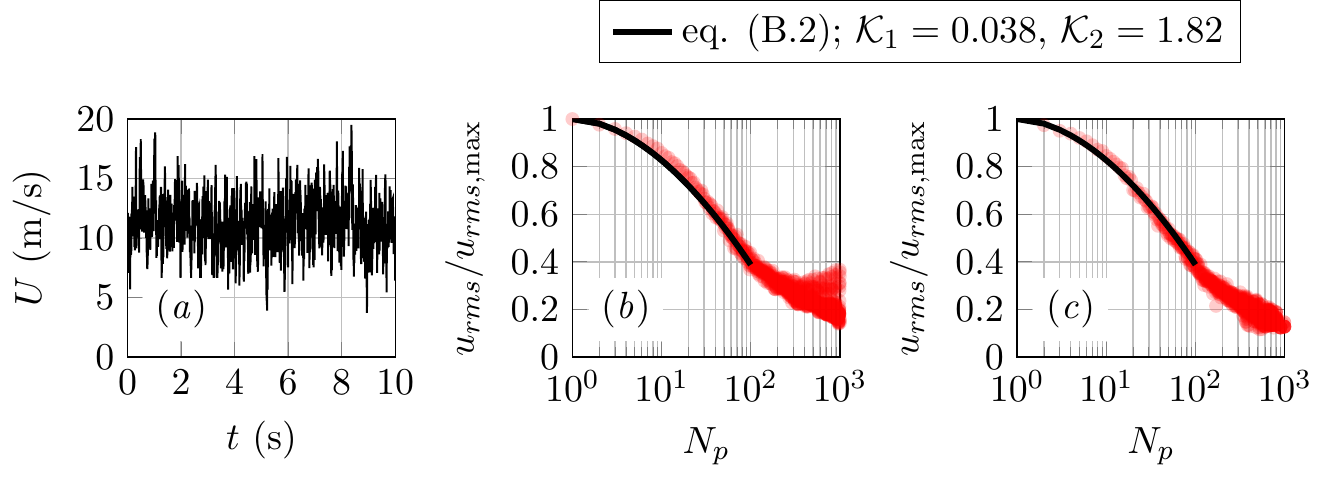}
\caption{Stochastic extrapolation of turbulent velocity fluctuations \textit{a}) time series of stochastic velocities $U$, generated using eq. (\ref{eq:stoch}) and $\langle U \rangle = 11$ m/s, $Tu$ = 0.2, $T_{uu} = 0.05$ s, $T = 45$ s (only 10 s shown for clarity) and $f = 200$ 1/s  (\textit{b}) normalised $rms$ fluctuations evaluated using different number of particles/measurement points from the stochastic signals shown on the left (\textit{c}) normalised $rms$ fluctuations for virtual phase-detection probe data from \cite{Kramer19AWCC} with $\langle U \rangle = 2.99$ m/s, $Tu$ = 0.2, $T_{uu} = 0.06$ s,  $C$ = 0.7, $F$ = 140 and $T = 90$ s.}
\label{fig:extrapolation}
\end{figure}

The signal was sampled at $f$ = 200 Hz for $\langle U \rangle$ = 11 m/s, $Tu$ = 0.2 and $T_{uu}$ = 0.05 s. The sampling duration was $T$ = 45 s and the root mean square of the velocity fluctuations was $u_{rms}(N_p=1) = u_{rms,\mathrm{max}}$ $\approx$ 2.2 m/s (figure \ref{fig:extrapolation}\textit{a}). In a second step, particle velocities were averaged over $N_p$ measured data points, followed by an evaluation of the $rms$ of the velocity fluctuations $(u_{rms}(N_p))$. The results (figure \ref{fig:extrapolation}\textit{b}) were approximated by a logarithmic expression: 
\begin{equation}
u_{rms}/u_{rms,\mathrm{max}} = 1 - \mathcal{K}_1  \ln (N_p)^{\mathcal{K}_2} \quad \quad \{ N_p \in \mathbb{Z}^{+}: 1 \leq N_p \leq 100 \}
\label{eq:fitnew}
\end{equation}
Eq. (\ref{eq:fitnew}) is valid up to $N_p$ $\approx$ 100 and the empirical constants $\mathcal{K}_1$ and $\mathcal{K}_2$ are a function of:
\begin{equation}
 \mathcal{K}_1,\mathcal{K}_2 = f_1\left(f,Tu,T_{uu}, ... \right)  
\end{equation}
The derived relationship was also confirmed using virtual phase-detection probe signals of synthetic particles from \cite{Kramer19AWCC} (figure \ref{fig:extrapolation}\textit{c}).

\end{document}